\RequirePackage{lineno}
\documentclass[a4paper,11pt]{article}
\pdfoutput=1
\usepackage{slashed}
\usepackage[affil-sl,auth-sc]{authblk}
\usepackage{amssymb,amsmath,amsbsy}
\usepackage{bbm}
\usepackage{mathrsfs}
\usepackage{mathbbol}
\usepackage{bm}
\usepackage{appendix}
\usepackage{hyperref}
\usepackage[top=1.2in, bottom=1.1in,left=0.9in,includefoot]{geometry}
\usepackage{subcaption}
\usepackage{graphicx}
\usepackage{cite}
\usepackage{float}
\usepackage{caption}
\usepackage{wasysym}
\usepackage{amsthm}
\usepackage[utf8x]{inputenc}
\usepackage[english]{babel}
\usepackage[capitalize]{cleveref}
\newif\ifContLineTwo
\newif\ifContLineThree
\def\conC#1{\vbox{\ialign{##\crcr
			\ifContLineThree\hrulefill\else\vphantom{\hrulefill}\fi\crcr
			\noalign{\kern3.2pt\nointerlineskip}
			\ifContLineTwo\hrulefill\else\vphantom{\hrulefill}\fi\crcr
			\noalign{\kern3.2pt\nointerlineskip}
			\ifContLineOne\hrulefill\else\vphantom{\hrulefill}\fi\crcr
			\noalign{\nointerlineskip}
			$\hfil\textstyle{\vbox to 14pt{}#1}\hfil$\crcr}}}
\def\DrawLeg#1#2{
	\kern-.2pt              
	\dimen2 =#1             
	\advance\dimen2 by 2pt  
	\dimen3 = 10.6pt        
	\dimen4 =3.6pt          
	\advance\dimen3 by -\dimen2 
	\multiply\dimen4 by #2
	\advance\dimen3 by \dimen4
	\raise\dimen2 \hbox{\vrule height\dimen3 width .4pt} 
	\kern-.2pt}             
\def\begC#1#2{\setbox0 =\hbox{$\textstyle{#2}$}
	\dimen0=.5\wd0 \dimen1=\ht0
	\conC{\hskip\dimen0}
	\count255=#1
	\ifnum\count255 =1 \ContLineOnetrue\else
	\ifnum\count255 =2 \ContLineTwotrue\else
	\ifnum\count255 =3 \ContLineThreetrue\fi\fi\fi
	\DrawLeg{\dimen1}{\count255}
	\conC{\hskip\dimen0}
	\kern-\dimen0\kern-\dimen0 \box0}
\def\endC#1#2{\setbox0 =\hbox{$\textstyle{#2}$}
	\dimen0=.5\wd0 \dimen1=\ht0
	\conC{\hskip\dimen0}
	\count255=#1
	\ifnum\count255 =1 \ContLineOnefalse\else
	\ifnum\count255 =2 \ContLineTwofalse\else
	\ifnum\count255 =3 \ContLineThreefalse\fi\fi\fi
	\DrawLeg{\dimen1}{\count255}
	\conC{\hskip\dimen0}
	\kern-\dimen0\kern-\dimen0 \box0}

\def\lsim{\buildrel{\scriptscriptstyle <}\over{\scriptscriptstyle\sim}}
\def\gsim{\buildrel{\scriptscriptstyle >}\over{\scriptscriptstyle\sim}}

\newcommand{\bea}{\begin{eqnarray}}
\newcommand{\eea}{\end{eqnarray}}

\title{\bf Probing mild-tempered neutralino dark matter through top-squark production at the LHC}
\author[a]{\small Monoranjan Guchait \thanks{guchait@tifr.res.in}~}
\author[a]{\small Arnab Roy \thanks{arnab.roy@tifr.res.in}~}
\author[b]{\small Seema Sharma \thanks{seema@iiserpune.ac.in}}
\affil[a]{\small Department of High Energy Physics,
	Tata Institute of Fundamental Research\\~~~~~~~~~~~~~~~~~~~~~~~~~~
	Homi Bhabha Road, Mumbai-400005, India}
\affil[b]{\small Indian Institute of Science Education and Research,
	Pune-411008,
	India}

\def \MET{E{\!\!\!/}_T}
\def\invfb{\text{fb}^{-1}}

\def\t {\widetilde {t_1}}
\def\N{\widetilde{\chi}^0}

\def\C1{\widetilde \chi_1^{\pm}}
\def\mst1 {m_{\t1}}
\def\br {\begin{eqnarray}}
\def\er {\end{eqnarray}}

\def\lsim{\buildrel{\scriptscriptstyle <}\over{\scriptscriptstyle\sim}}
\def\gsim{\buildrel{\scriptscriptstyle >}\over{\scriptscriptstyle\sim}}
\def \MET{E{\!\!\!/}_T}
\def\invfb{\text{fb}^{-1}}

\date{}

\begin{document}

		\maketitle
	\begin{abstract}
		The lightest neutralino, assumed to be the lightest supersymmetric particle, is proposed to be a dark matter (DM) candidate for the mass $\cal{O}$(100) GeV. Constraints from various direct
		dark matter detection experiments and Planck measurements exclude a substantial
		region of parameter space of minimal supersymmetric standard model (MSSM). 
		However, a ``mild-tempered'' neutralino with dominant bino composition
		and a little admixture of Higgsino is found to be a viable candidate for DM.
		Within the MSSM framework, we revisit the allowed
		region of parameter space that is consistent with all existing constraints. Regions of parameters that are not sensitive to
		direct detection experiments, known as ``blind spots,'' are also revisited.
		Complimentary to the direct detection of DM particles, a
		mild-tempered neutralino scenario is explored
		at the LHC with the center of mass energy $\rm \sqrt{s}$=13~TeV through the top-squark pair production, and its subsequent decays
		with the standard-model-like Higgs boson in the final state. Our considered channel
		is found to be very sensitive also to the blind spot scenario. 
		Detectable signal sensitivities are achieved using the cut-based method
		for the high luminosity options $\rm 300$ and $\rm 3000 ~fb^{-1}$,
		which are further improved by applying the multi-variate analysis
		technique. 
	\end{abstract}
	\vskip .5 true cm
\newpage

	\section{Introduction}
	The quest for a signature of beyond standard model (SM) physics
	is a very high priority agenda in high energy physics
	experiments and it has been going on for a long time in
	several laboratories. In particular, at the LHC experiments,
	looking for new physics signals is the major thrust area.
	Unfortunately, no single direct evidence of new physics
	signals has been observed at this point. As a consequence, the absence of
	experimental confirmation leads to stringent constraints to
	various BSM models\cite{Rappoccio:2018qxp}. 
	On the other hand, the well-confirmed existence of dark matter (DM)
	by various cosmological and astrophysical experiments serves as one of the
	strong motivations to propose the existence of BSM physics\cite{Bertone:2004pz,Feng:2010gw}.
	Among several probable candidates of DM, the weakly interacting massive
	particle (WIMP) turns out to be the most suitable one for thermal DM,
	with a correct relic density measured
	by the PLANCK experiment which predicts\cite{Aghanim:2018eyx}, 
	\br
	\rm \Omega h^2=0.12~\pm~0.001.
	\label{eq:wmap}
	\er
	Enormous efforts have been in place for a long time to look for DM candidates
	via direct and indirect searches in various experiments\cite{Schumann:2019eaa,Gaskins:2016cha,Slatyer:2017sev,Boveia:2018yeb,universe4110131,Aaboud:2019yqu}.
	However, null results, in particular, from some of the direct
	detection (DD) experiments have resulted in
	strong constraints on DM-nucleon scattering cross sections
	in terms of DM (WIMP) masses\cite{Akerib:2016vxi,Cui:2017nnn,Aprile:2018dbl,Agnes:2018ves,Agnese:2018col,Aprile:2019dbj,Amole:2019fdf,Adhikari:2019off,Ajaj:2019imk,Abdelhameed:2019hmk}.
	The DM-nucleon scattering cross section can be classified into two
	categories, namely, spin-independent (SI) and spin-dependent (SD),
	depending on the structure of the coupling. Note that, in general, the SI DM-nucleon scattering cross section is smaller than that of the SD case, and it is more sensitive to DD experiments\cite{Barger:2008qd,Belanger:2008sj,Agrawal:2010fh}.
	For instance, the most stringent bounds come from XENON1T experiment, where
	the DM-nucleon scattering cross section corresponding to the DM of
	the mass range $\sim$20--100 GeV is strongly 
	restricted, $\rm \sigma_{SI}\lsim10^{-46} {\rm cm}^2$\cite{Aprile:2018dbl}. The other experiments
	such as LUX\cite{Akerib:2016vxi}, PANDA\cite{Cui:2017nnn}, PICO-60\cite{Amole:2019fdf}, Darkside\cite{Agnes:2018ves} etc. also constrain the DM-nucleon cross section for
	a wide range of masses of DM candidates from few GeV to TeV.
	
	The minimal supersymmetric standard model (MSSM) with R-parity conservation
	offers the lightest neutralino, assumed to be the lightest supersymmetric particle (LSP) as
	the potential DM (WIMP) candidate
	of the mass $\sim$ 100 GeV\cite{Goldberg:1983nd,Ellis:1983ew,Drees:1992am,Drees:2018dsj}. Comprehensive searches of neutralino DM are carried out at the LHC which lead to various constraints in the absence of any signal\cite{Belanger:2013pna,Choudhury:2013jpa,Han:2014nba}. In MSSM, the physical neutralino
	state is constituted through the linear superposition
	of electroweak (EW) gauginos (bino($\rm \widetilde B$), wino($\rm \widetilde W$))
	and Higgsinos($\rm \tilde H_u^0,\tilde H_d^0$). This composition is
	mainly determined by relative values of two EW gaugino mass parameters,
	$\rm M_1$ and $\rm M_2$, corresponding to U(1) and SU(2) gauge
	transformations respectively. In addition, the other two parameters, namely
	Higgsino mass parameter($\mu$) and $\tan\beta$, the ratio of two vacuum
	expectation values of two neutral Higgs bosons also play an important role
	in determining the physical masses and composition of neutralino states.
	A neutralino state with pure Higgsino or wino composition of the mass
	$\rm \sim\mathcal{O}(100)~GeV$ is found not to be a favorable DM
	candidate because of its under-abundance of relic density\cite{ArkaniHamed:2006mb}.
	However, for large masses $\rm\mathcal{O}(TeV)$, those can serve as a 
	DM candidate\cite{ArkaniHamed:2006mb,Chakraborti:2017dpu,Cahill-Rowley:2014boa,Chakraborti:2014fha,Delgado:2020url}.
	Similarly, a neutralino with pure bino composition also does
	not satisfy the right relic density measurement (see Eq.~\ref{eq:wmap}).
	Hence, in order to propose LSP as a viable DM candidate,
	the ``tempered neutralino'' scenario is proposed to be the best bet 
	\cite{ArkaniHamed:2006mb},
	where the neutralino is no longer a pure state, but has
	admixtures of more than one composition.
	A well-tempered bino-Higgsino\cite{Pierce:2004mk,Gogoladze:2010ch,vanBeekveld:2016hug,Abdughani:2017dqs} or bino-wino\cite{Baer:2005zc,Baer:2005jq,BirkedalHansen:2001is} neutralino
	is found to be the most suitable DM candidate for the mass
	$\sim$100~GeV to achieve the right relic density.
	The Higgsino component is indispensable to bring down the
	relic density to the required value (Eq.~\ref{eq:wmap}) via resonant Z or Higgs-mediated
	annihilation, where the Higgs can be the SM-like Higgs boson as well as
	heavier Higgs boson states, in the  limit  of  large  sfermion  masses.
	It is to be noted that, the neutralino-nucleon SI
	scattering cross section is enhanced with the increase of
	Higgsino composition in the neutralino state.
	Therefore, the strong experimental limits on the SI scattering cross section
	restrict the composition of neutralino, in particular, the Higgsino content\cite{Baer:2016ucr,Badziak:2017the}.
	Hence a bino dominated neutralino with a little mixture of Higgsino
	component, referred to as ``mild-tempered neutralino'', is
	expected to be the viable DM candidate for the
	mass $\rm \mathcal{O}(100)~GeV$ or little less~\cite{Profumo:2017ntc},
	and consistent with all existing constraints.
	In this regard, it is to be noted that few studies exist in the
	literature based on the extended supersymmetric (SUSY) model, which present very light DM candidates
	($\rm m_{\chi}\lsim50~GeV$)\cite{Guchait:2020wqn,Abdallah:2019znp,Barman:2020vzm,Barman:2020zpz,Abdallah:2020yag},
	satisfying all current constraints.
	
	It is worth pointing out here, that there exists
	a region of MSSM parameter space where the DM-nucleon SI scattering cross section almost vanishes because of the interplay among various amplitudes.
	Consequently, direct detection rate of DM becomes insensitive corresponding to that region of
	parameter space, which is known as the ``blind spot''(BS)  
	\cite{Cheung:2012qy,Cheung:2013dua,Huang:2014xua,Han:2016qtc,Cao:2019qng}.
	As the DD experiments fail to probe this BS scenario, it is worth finding a complementary way for DM searches at the LHC.
	
	In this current study, we focus on the mild-tempered scenario, i.e. bino-Higgsino neutralino with a larger bino component, and
	of the mass ${\cal O}$(100)~GeV, and then identify the corresponding
	region of parameter space consistent with all measurements. The existence of a relatively lighter LSP of the mass range considered in this study is still not absolutely ruled out by any SUSY searches at the LHC. Hence, our study will presumably give some idea about its detectability at the LHC with its high luminosity options.  
	With this aim, the characteristic signature corresponding to this mild-tempered neutralino
	including the BS scenario are discussed for the LHC experiment. We consider the top-squark pair production and then its cascade decay to SM-like Higgs boson and an LSP, the DM candidate. Although $\rm m_{\t}\lsim 1.1~TeV$ are ruled out from searches at the LHC in the context of various simplified models, for low $\rm BR(\t\rightarrow\N_1 +t)\sim 10\%$, $\rm m_{\t}<1~TeV$ are found to be still allowed using statistical analysis. It is be noted that the top-squarks of lower mass range which are within the reach of current LHC energy, are also motivated in the context of ``naturalness" scenario~\cite{Barbieri:1987fn,Giusti:1998gz,Kitano:2006gv,Barbieri:2009ev,Asano:2010ut,Baer:2012cf}. A detailed investigation is carried out performing simulation to explore the feasibility of finding the signal at the LHC for higher luminosity options, such as 
	$\rm {\cal L}=300 ~fb^{-1}$ and 3000~fb$^{-1}$. 
	
	The paper is organized as follows. In section 2, the MSSM model set up providing mild-tempered neutralino and BS scenario is discussed, and then corresponding allowed region of parameters are identified. In section 3, signal and background simulations are presented and followed by results. Finally, we summarize in section 4.
	
	\section{Mild-tempered neutralino scenario in the MSSM}
	
	In this section, we discuss the MSSM model setup and then delineate the
	region of parameter space interesting to our scenario which presents
	a DM candidate of mass $\sim {\cal O}$(100)~GeV consistent
	with the existing data from Planck experiment (Eq.~\ref{eq:wmap}) and direct searches as mentioned
	above.
	
	In the gauge eigenstate basis ($\rm \widetilde{B}, \widetilde{W}_3, \widetilde{H}_d^0, \widetilde{H}_u^0$),
	the neutralino mass matrix can be written as,
	\br
	\rm {M}_N = \left( \begin{array}{cccc}
		\rm M_1 &0 &\frac{-g_1 v c_{\beta}}{\sqrt 2} & \frac{g_1 v s_{\beta}}{\sqrt 2}  \\
		0 &\rm M_2   & \frac{g_2 v c_{\beta}}{\sqrt 2} & \frac{-g_2 v s_{\beta}}{\sqrt 2} \\
		\frac{-g_1 v c_{\beta}}{\sqrt 2} & \frac{g_2 v c_{\beta}}{\sqrt 2}  & 0 & - \mu \\
		\frac{g_1 v s_{\beta}}{\sqrt 2}  & \frac{-g_2 v s_{\beta}}{\sqrt 2}  &  -\mu  & 0
	\end{array} \right).
	\label{eq:mneu}
	\er
	Here,  $\rm M_1(g_1)$ and $\rm M_2(g_2)$ present the (U(1))$\rm \widetilde B$ and 
	(SU(2))$\rm \widetilde W_3$ gaugino mass(coupling) parameters respectively, whereas $\mu$ is defined to be the Higgsino mass parameter. The two VEVs corresponding to two neutral components of the two Higgs doublets $\rm H_u^0$ and $\rm H_d^0$ are $v_u$ and $v_d$ respectively
	and constrained to be $v_u^2 + v_d^2=v^2$. As practice, we assume
	$\tan\beta=\frac{v_u}{v_d}$, and $\rm s_{\beta}\equiv sin \beta,\; c_{\beta}\equiv cos\beta$. The symmetric matrix $\rm {M_N}$ can be diagonalized by a unitary matrix $\rm N_{4\times 4}$ to obtain the masses of four neutralino states $\rm \tilde{\chi}_i^0 (i=1,2,3,4)$ as,
	\br
	\rm M_{\tilde \chi^0}^D  = \rm N M_N N^{\dagger},
	\label{eq:nn}
	\er    
	and the corresponding physical neutralino states are given by, 
	\begin{equation}
	\rm \widetilde{\chi}_i^0=N_{i1}{\widetilde B}+N_{i2}{\widetilde W_3}+
	N_{i3}\widetilde{H}_d^0+N_{i4}\widetilde{H}_u^0.
	\end{equation}	
	Among the four neutralino states, two of the lighter states become gaugino-like ($\rm \widetilde{B}$ and $\widetilde{W}_3$), if $\rm |M_{1,2}-\mu|\geq M_Z$ and $\rm |\mu|> M_2>M_1$ with masses $\rm m_{\N_1}\sim M_1$ and $\rm m_{\N_2}\sim M_2$ respectively. The masses of Higgsino dominated states are  mostly controlled by $\mu$, and in particular $\rm m_{\N_{2,3}}\sim \mu$ for a decoupled
	scenario ($\rm M_2 >> \rm \mu>M_1$). Further, for $\rm M_1<\mu<<M_2$ cases,
	the heaviest state is expected to be $\rm \sim \widetilde W_3$-like,
	whereas intermediate states
	become Higgsino dominated with the lightest state almost bino-like
	with tiny Higgsino component ($\rm i.e.,~ N_{11}^2>>N_{13}^2+N_{14}^2$).  	
	Similarly, in the basis ($\rm i\widetilde{W}^-, \widetilde{H}_u^-$) and ($\rm i\widetilde{W}^+, \widetilde{H}_d^+$) the chargino mass matrix is given by:
	\begin{equation}
	\rm M_C = \left(\begin{array}{c c}
	\rm M_2 & \rm \sqrt{2} M_W \sin\beta\\
	\rm \sqrt{2} M_W \cos\beta & \mu
	\end{array}
	\right),
	\end{equation}
	which is diagonalized by two unitary matrices U and V. For $\rm M_2>>\mu$, the lighter chargino ($\widetilde{\chi}_1^{\pm}$) state becomes Higgsino-like.
	
	For our considered scenario, the dominant DM annihilation process
	occurs through the s-channel mediated by CP-even(h, H) and CP-odd(A)
	Higgs bosons or Z, 
	\br
	\rm \N_1\N_1\xrightarrow{{\phi/Z}}f\bar{f},\ \ \ \ \phi=h,H,A, 
	\label{eq:hannihilation}
	\er
	in the limit of relatively heavier slepton masses.
	The cross section of the annihilation process primarily depends on the 
	$\rm (\phi,Z)$-$\N_1$-$\N_1$ couplings, which are of the following form,
	\br
	\rm g_{h\N_1\N_1} &\sim& \rm g(N_{12}-\tan\theta_WN_{11})(sin\alpha N_{13}+\cos\alpha N_{14})
	\label{eq:hchichi}\\
	\rm g_{H\N_1\N_1} &\sim& \rm g(N_{12}-\tan\theta_WN_{11})(sin\alpha N_{14}-\cos\alpha N_{13})
	\label{eq:Hchichi}\\
	\rm g_{A\N_1\N_1} &\sim& \rm g(N_{12}-\tan\theta_WN_{11})(\cos\beta N_{14}-sin\beta N_{13})
	\label{eq:achichi}\\
	\rm g_{Z\N_1\N_1} &\sim& \rm \frac{g}{2\cos\theta_W}(N_{13}^2-N_{14}^2),
	\label{eq:higgs-couplings}
	\er
	where $\rm \alpha$ is the mixing angle of the CP-even Higgs sector.
	Clearly, the combined effect of bino$\rm (N_{11})$ and
	Higgsino components($\rm N_{13},N_{14}$) 
	in $\rm \N_1$ determine the annihilation rate.
	In addition, a neutralino with a moderate to large amount of Higgsino
	content may dominantly co-annihilate with Higgsino-like(large $\rm V_{12}$) and nearly mass degenerate
	lighter chargino $\rm \widetilde{\chi}_1^{\pm}$,
	\br
	\rm \N_1\widetilde{\chi}_1^{\pm}\xrightarrow{{W^{\pm}}}f\bar{f},
	\label{wcoanni}
	\er
	along with other subdominant contributions, which may enhance the annihilation cross-section through
	the following coupling,
	\br 
	\rm g_{W^\pm\N_1\widetilde{\chi}_1^{\pm}}=\frac{g\tan\theta_W}{\sqrt{2}}(N_{14}V_{12}^*-\sqrt{2}N_{12}V_{11}^*).
	\er
	Hence, in combination of all processes, whichever are viable, the cross section for annihilation process corresponding to a Higgsino-like LSP goes up leading to
	an under-abundance of relic density.
	Hence one can conclude that an LSP with a suitable combination of bino
	and Higgsino composition appears to be a viable DM candidate around the
	mass ${\cal O}$(100)~GeV. In the case of wino-Higgsino dominated
	LSP, various possible annihilation and co-annihilation processes can
	occur, which are mediated by SM gauge bosons, and lead to the under-abundant scenario. In order to achieve the right relic density prediction, in this case, one needs to lift the mass($\rm \sim M_2$) of the LSP to TeV level and suppress annihilation cross section\cite{ArkaniHamed:2006mb,Chakraborti:2017dpu}. This type of scenario appears naturally in anomaly mediated SUSY breaking model\cite{Moroi:1999zb,Gherghetta:1999sw,Arbey:2011gu}.
	Hence, a SUSY DM model disfavors the possibility of
	Higgsino/wino dominated scenario with DM mass $\rm\sim \cal{O}$(100) GeV.
	
	As pointed out earlier, the
	composition of the LSP DM candidate is also constrained
	by direct detection experiments\cite{Aprile:2018dbl,Akerib:2016vxi,Cui:2017nnn,Aprile:2019dbj,Amole:2019fdf},
	where DM candidate scattering off a heavy nucleus mediated by Higgs/gauge
	bosons or squarks. The effects due to heavier 
	squarks (${\cal O}$(1) TeV) are very much suppressed. Hence the main contribution
	to SI(SD) cross section occurs through Higgs(Z) boson exchange via t-channel
	diagram\cite{Barger:2008qd,Belanger:2008sj}. The dominant contribution 
	comes from the diagram mediated by the CP-even lightest Higgs boson, 
	whereas contributions due to other heavier Higgs bosons are suppressed. 
	Interestingly, this suppression can be
	compensated by enhanced couplings of heavier Higgs bosons with the quarks
	for a certain range of parameters, in particular, for higher values of
	$\tan\beta$, which we will discuss later.
	
\begin{figure}[H]
		\begin{subfigure}[b]{0.49\textwidth}
			\centering
			\includegraphics[width=6.9cm]{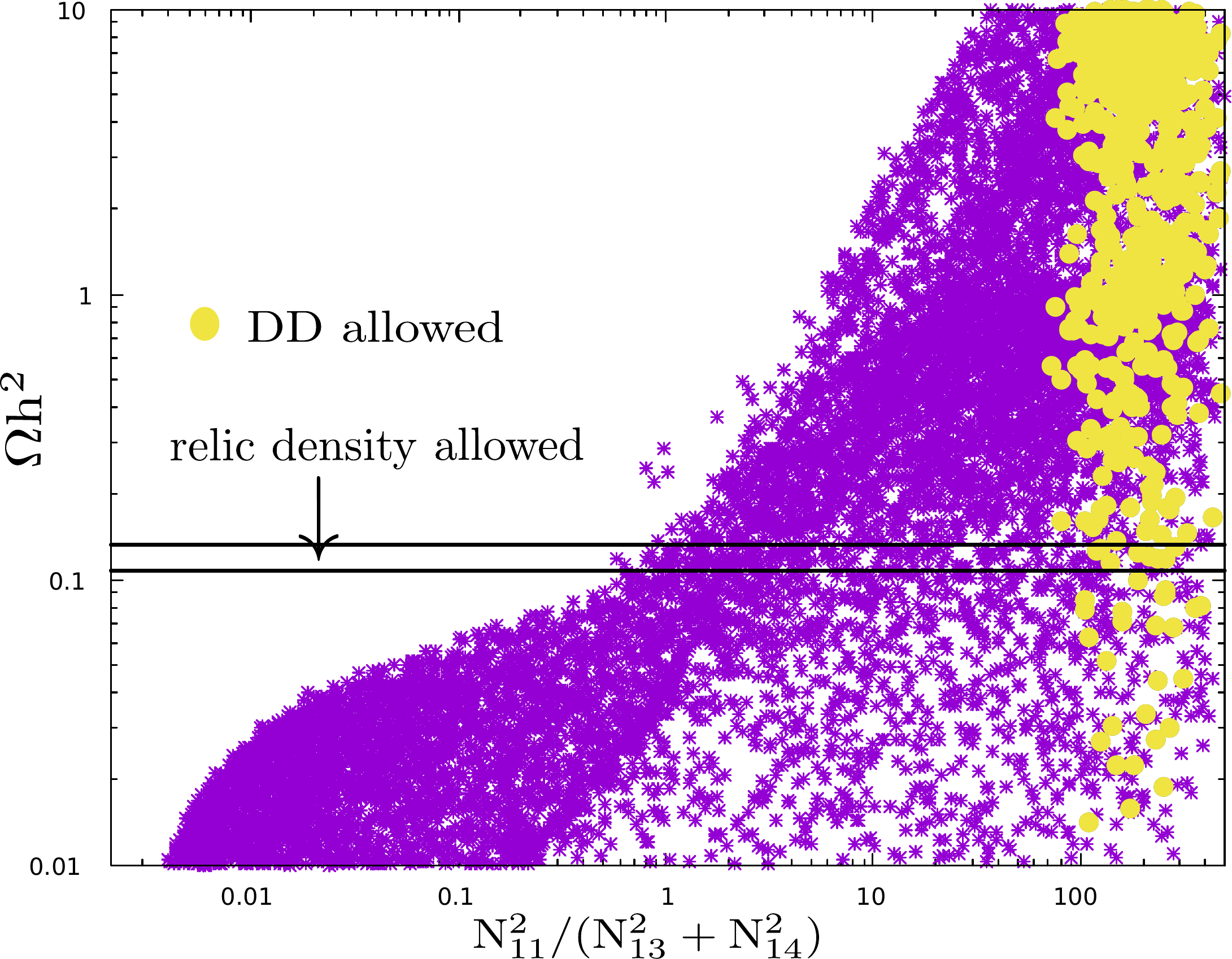}
		\end{subfigure}
		\begin{subfigure}[b]{0.49\textwidth}
			\centering
			\includegraphics[width=7.5cm]{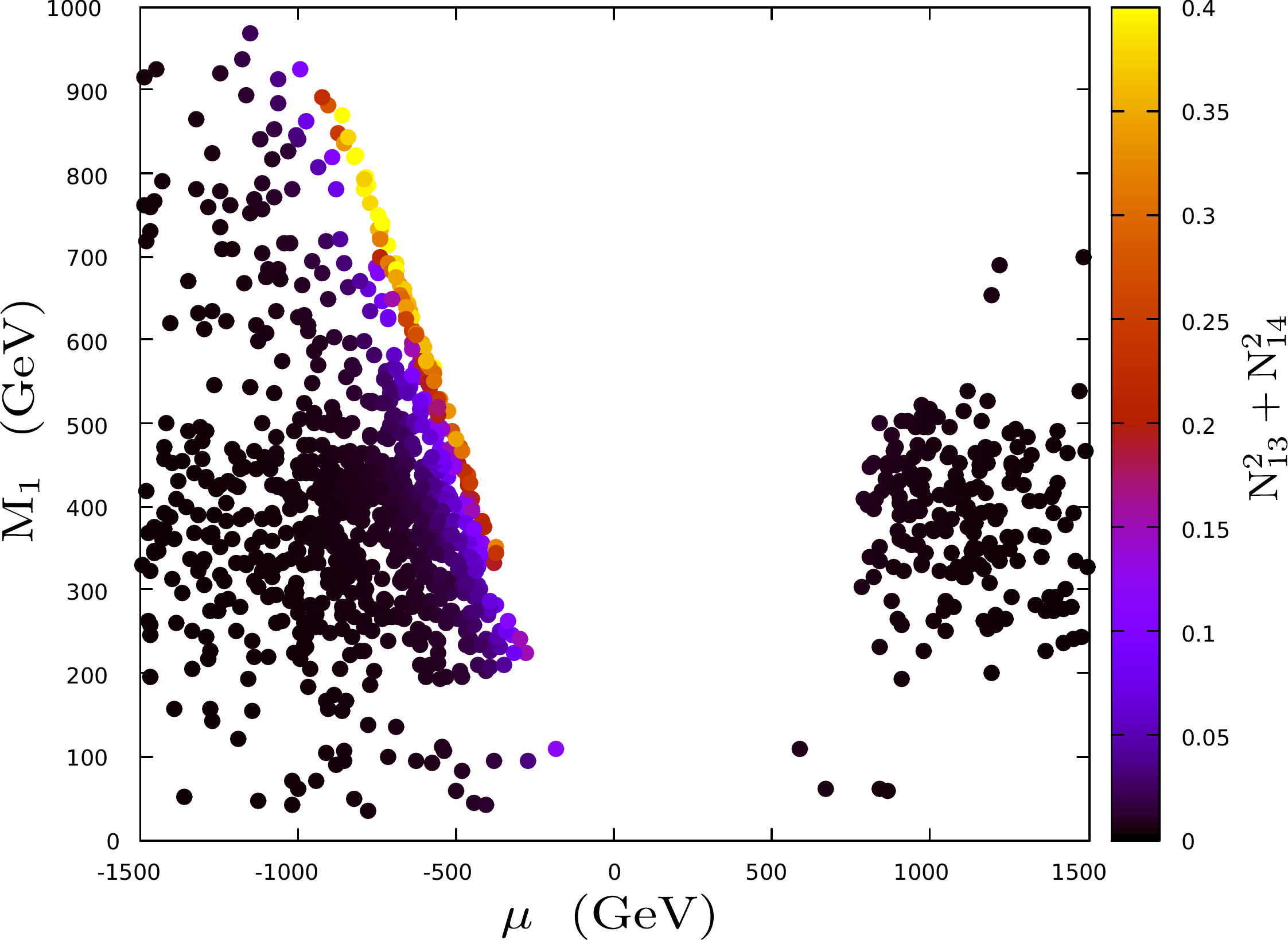}
		\end{subfigure}
		\caption{\small (left) Relic density with the variation of $\rm \tilde{B}/\tilde{H}$ components in $\rm \N_1$. Allowed points by DD measurements(yellow) and relic density(black band); (right) Ranges of $\mu$ and $\rm M_1$ along with the Higgsino component of the LSP allowed by relic density and DD constraints.}
		\label{fig:relic}
	\end{figure}

	The SI scattering cross section is also
	sensitive to couplings $\rm g_{h\N_1\N_1}$(Eq.~\ref{eq:hchichi}). 
	The presence of a larger Higgsino component in $\rm \N_1$     
	enhances the SI DM-nucleon scattering cross section mediated 
	mainly by the CP even lightest Higgs boson, which is tightly constrained
	by the existing limits on DM-nucleon 
	scattering cross section from the XENON1T experiment\cite{Aprile:2018dbl}.
	On the other hand, the composition of the $\rm \N_1$ state is also constrained by relic density.  
	Hence, the bino-Higgsino content of a mild-tempered neutralino state is
	severely restricted by the combined effect of relic density measurements and
	DM-nucleon cross section limits.
	This feature of mild-tempered neutralino is reflected in Fig.~\ref{fig:relic}(left), where the variation of relic density
	with the relative size of bino and Higgsino composition of the LSP is presented in terms of $\rm N_{11}^2/(N_{13}^2 + N_{14}^2)$. It is to be noted that this figure is
	subject to the condition $\mu>0$ to avoid effects from ``blind spots,'' which occur for $\mu<0$ when $\rm M_1$ is assumed to be positive. It will be discussed in detail later. In Fig.~1(right), we present the ranges of $\rm \mu$ and $\rm M_1$ allowed by relic density, limits from DD experiments along with some other constraints as described in section 2.1.
	These figures are obtained by performing a numerical scan of parameters (Eq.\ref{eq:nmssmpara}), which will be discussed later.
	Fig.~\ref{fig:relic}(left) indicates that the $\rm \N_1$, with relatively higher Higgsino
	composition and tiny bino content, makes under-abundance of relic density.
	In this case, along with DM annihilation (Eq.~\ref{eq:hannihilation}), the co-annihilation
	process(Eq.~\ref{wcoanni}) also takes place resulting in a larger
	DM annihilation cross section. Towards the rightmost region of Fig.~\ref{fig:relic}(left), due to the absence of sufficient Higgsino-components in the LSP, the DM-nucleon scattering cross section goes down
	because of the couplings(Eq.~\ref{eq:hchichi}) and becomes
	consistent with DD limits. This region is 
	presented(yellow) in Fig.~\ref{fig:relic}(left) at the higher values of ratio
	$\rm N_{11}^2/(N_{13}^2 + N_{14}^2)$. Hence, it can be concluded that the bino dominated
	LSP with little admixtures ($\sim 1\%$) of Higgsino component 
	is the most favoured option in a decoupled scenario ($\rm M_2$ is very large). We referred to this as a scenario of ``mild-tempered'' neutralino in the previous section.
	In this scenario, relatively higher values $\rm \mu$ are found to be allowed with light to moderate values of $\rm M_1$. It is clearly seen in Fig.~1(right) that, for $\mu>0$, the Higgsino fraction in the LSP is tiny. However, for $\mu<0$, there exist parameter spaces with a comparatively higher amount of Higgsino components that are still allowed. It occurs mainly due to the effect of ``blind spots'', which is discussed next.
	
	The blind spot is an interesting scenario where the 
	DM-nucleon scattering cross section is found to be very  
	insensitive for a certain range of relevant parameters in the MSSM, and the
	corresponding region of parameters is called ``blind spot''. It may
	happen for various reasons. For instance, the tree-level scattering
	cross section may vanish either for a pure gaugino 
	(i.e $\rm N_{13}, N_{14} \sim 0$) or Higgsino (i.e $\rm N_{11} \sim 0$)
	neutralino state.
	Moreover, scattering takes place via one and two loop diagrams
	mediated by gauge bosons, and an accidental cancellation 
	among various scattering amplitudes for pure Higgsino and 
	gaugino LSP state, lead the total cross sections
	too small and beyond the sensitivity of DD
	experiment( $\rm \sigma_{SI} << 10^{-46} {\rm cm}^2$)\cite{Hisano:2011cs,Hill:2011be,Cheung:2012qy}.
	Finally, BS may also arise at the tree level due to cancellation among
	various amplitudes. 
	The dominant contribution to DM-nucleon
	cross section comes from the diagram mediated by the CP
	even lightest Higgs boson, whereas contributions due to other heavier
	Higgs bosons are found to be very small for the 
	decoupling scenario($\rm m_A >> M_Z$).
	Interestingly, at the tree level, the suppression of contribution mediated 
	by heavier Higgs bosons can be
	compensated by its enhanced coupling with the 
	(down type) fermions for the range of moderate
	to higher values of $\tan\beta$.
	Additionally, the coupling between heavier Higgs bosons and 
	neutralinos, $\rm H$-$\N_1$-$\N_1$(Eq.~\ref{eq:Hchichi}) may receive
	similar kind of enhancement for a larger value of $\rm N_{13}$,
	the down type of Higgsino content in the LSP. 
	Consequently, the amplitudes mediated by heavier 
	Higgs bosons turn out to be comparable or at the same level 
	of the CP even SM-like Higgs boson exchange diagram.
	Depending on the relative signs of $\mu$ and $\rm M_{1}$,
	the interference between these two diagrams,
	may become destructive or constructive\cite{Huang:2014xua}.       
	Incidentally, for a certain range and combination of related
	parameters, these two contributions almost cancel each other 
	leading to the scattering cross section insensitive\cite{Huang:2014xua}. 
	A detailed analytical study shows that 
	the combination of parameters corresponding 
	to the BS for moderate to larger values of $\tan\beta$ 
	follow the relation among  
	$\mu$, $\rm m_{A}$, $\rm tan\beta$, and $\rm m_{\N_{1}}$ ($\rm \sim M_1$),
	as~\cite{Huang:2014xua},
	\br
	\rm \frac{M_1}{\mu}\sim -\left(\sin 2\beta+tan\beta\frac{m_h^2}{2m_A^2}\right).
	\label{eq:bseq}
	\er
	Corresponding to this parameter space, naturally a larger 
	Higgsino component $\sim \cal{O}$$(10\%)$ can be accessible without 
	violating DD bounds in contrast to the requirement of $\sim \cal{O}$$(1\%)$ 
	or less for a mild-tempered neutralino case.
	The above condition for BS connects the gaugino mass parameter with the Higgs
	sector.
	\begin{figure}[H]
			\centering
			\includegraphics[width=8 cm]{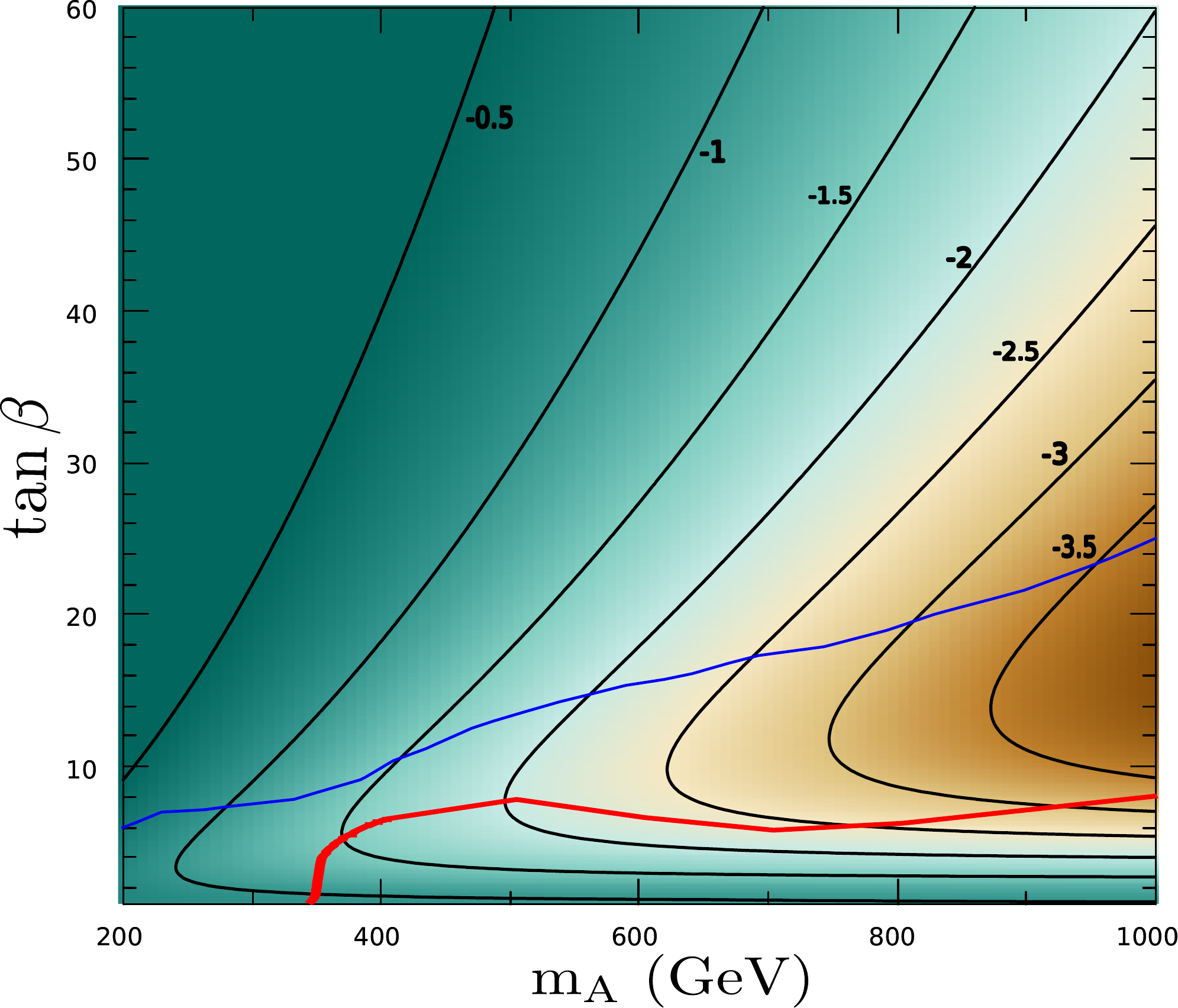}
		\caption{Contour plots of $\rm \mu/m_{\N_{1}}$ for various values corresponding to BS scenario in the $\rm tan\beta$ and $\rm m_A$ plane along with the exclusion lines from $\rm h,A \to \tau\tau$ searches in ATLAS(red) and CMS(blue) experiments.}
		\label{fig:BS}
	\end{figure}
	Note that a substantial region of $\rm \tan\beta$ and
	$\rm m_A$ plane is excluded from the Higgs searches in the channel,
	$\rm h,A \to \tau\tau$ \cite{Aad:2020zxo,Sirunyan:2018zut}.
	This $\rm m_A$-$\tan\beta$ exclusion can be traded to obtain constraints
	on $\rm \mu/m_{\N_1}$, by using Eq.~\ref{eq:bseq} in the $\rm m_A$-$\tan\beta$
	plane. In Fig.~\ref{fig:BS}, following Eq.~\ref{eq:bseq}
	the contour plots of $\rm \frac{\mu}{m_{\N_1}}\sim \frac{\mu}{M_1}$
	are shown in the $\rm m_{A}$, $\rm tan\beta$ plane\cite{Huang:2014xua}.
	The region above the red and blue lines are excluded due to the non observation of any signal events in the $\rm h,A \to \tau\tau$
	searches by ATLAS\cite{Aad:2020zxo}($\rm \mathcal{L}= 139.5~fb^{-1}$) and CMS\cite{Sirunyan:2018zut}($\rm \mathcal{L}= 35.9~fb^{-1}$) experiments respectively. Depending on the value of $\rm m_A$, the
	BS condition, i.e. the ratio $\rm \frac{\mu}{M_1}$ may vary from -1.5 to -3.5.
	It implies that the lightest neutralino state is bino like whereas the second and third
	heavier states are Higgsino like in the limit of large $\rm M_2$,
	which is exactly the scenario that we try to explore at the LHC experiment.
	
	As explained before, the main focus of this study is to explore the
	feasibility of finding mild-tempered neutralino scenario at the LHC.
	The added advantage of our proposed channel is its sensitiveness
	to the region of parameters corresponding to the BS scenario,
	which can also be probed at the LHC.
	As we know, the content of bino and Higgsino in the LSP
	depends on the splitting between $\rm \mu$ and $\rm M_1$. Therefore,
	mild-tempered scenario appears with the condition $\rm |\mu|-M_1\gsim 100~GeV$, which provides also
	an LSP of mass $\cal O$(100) GeV.
	The scenario with little   
	Higgsino admixture along with the dominant bino composition   
	in $\rm \N_1$ presumably predicts  
	Higgsino-dominated $\N_{2,3}$ and $\widetilde{\chi}_1^{\pm}$ states
	that are degenerate in mass $\sim \mu$, for a decoupled
	wino state (i.e large $\rm M_2$).
	In such cases, heavier states $\N_{2,3}$ prefer to
	decay to a Z boson and an LSP, and  
	$\widetilde{\chi}_1^\pm$ decays to a W and an LSP. The coupling involved in
	$\N_{2,3}$ decays is   
	$\rm Z$-$\N_{2,3}$-$\N_1 \propto $ $\rm N_{13} N_{23} - N_{14} N_{24}$, and
	since $\N_1$ is primarily bino dominated (i.e $\rm N_{13},N_{14}$ very tiny),
 it is suppressed.
	Thus $\rm \N_{2,3}$
	preferably decay as,
	\br
	\rm \N_{2,3} \to h + \N_1,
	\er
	and a larger Higgsino composition in $\N_{2,3}$ state(Eq.~\ref{eq:hchichi})
	makes its rate higher.
	This decay channel of $\N_{2,3}$ is found to be the 
	characteristic feature for the mild-tempered neutralino scenario. 
	Hence testing of this scenario can be performed by studying
	$\N_{2,3}$ and $\widetilde{\chi}_1^\pm$
	production at the LHC\cite{Gori:2011hj,Ghosh:2012mc} and their subsequent decays.
	Earlier, this channel is thought to be a ``spoiler'' mode
	corresponding to trilepton signal in
	pp$\to \N_2\widetilde{\chi}_1^\pm\to \ell^+ \ell^- \ell \N_1\N_1$
	production\cite{Barger:1994nm,Baer:1994nr}.
	In this study, instead of considering the $\rm \N_2 \widetilde{\chi}_1^\pm$ 
	production via electroweak interaction, we consider the production
	of $\N_2$ through lighter top-squark production via strong interaction, where     
	$\rm \t$ dominantly decays to Higgsino-like $\rm \N_{2,3}$ 
	and $\chi_1^{\pm}$\cite{Dutta:2013sta,Ghosh:2012mc}. The decay $\rm \t \to t +\N_{2,3}$,   
	is governed by the interactions, 
	\br
	\rm \mathcal{L}_{t\bar{\tilde{t}}\tilde{\chi}_i^0} = \bar{t}\left(g_{L}^{\tilde{\chi}_i^0}P_L+g_{R}^{\tilde{\chi}_i^0}P_R\right)\N_{i}\tilde{t},
	\label{eq:ffchi}
	\er
	where,
	\br
	g_{L}^{\tilde{\chi}_i^0} &=& -\left[\frac{g_2}{\sqrt{2}}N_{i2}+\frac{g_1}{3\sqrt{2}}N_{i1}\right]\cos\theta_{\tilde{t}} - \frac{m_t}{v} N_{i4}\sin\theta_{\tilde{t}}\nonumber\\
	&\sim& \frac{g_1}{3\sqrt{2}}N_{i1}\cos\theta_{\tilde{t}} - \frac{m_t}{v} N_{i4}\sin\theta_{\tilde{t}}\\
	g_{R}^{\tilde{\chi}_i^0}&=&\frac{2\sqrt{2}}{3}g_1N_{i1}\sin\theta_{\tilde{t}}-\frac{m_t}{v}N_{i4}\cos\theta_{\tilde{t}}.
	\er
	Here $\theta_{\tilde t}$ is the mixing angle in the top-squark sector. 
	Evidently, the $\rm m_t$ dependent term becomes dominant
	for Higgsino($\rm N_{i4}$)-like neutralino states
	leading higher branching ratio (BR) for $\rm \t  \to \N_{2,3} + t$. 
	Due to large enough splitting between 
	$\mu$ and $\rm M_1$, it is natural to have $\rm m_{\N_{2,3}}-m_{\N_1}>125 ~GeV$, 
	resulting in the $\rm \N_{2,3}\rightarrow h +\N_1$ decay to be 
	dominant.	
	Hence the mild-tempered neutralino DM can be indirectly 
	produced from the decay of Higgsino-like $\rm \N_{2,3}$ 
	producing those in the lighter top-squark($\rm \t$) production.
	It is be noted that, this type of scenario can also be probed through the associated
	production, such as $\rm p p \to \N_{2,3}\widetilde{\chi}_1^\pm$ and with the three lepton final
	states along with missing energy\cite{Han:2014sya,vanBeekveld:2016hbo,Liu:2020muv}. However, we observe that for the
	same set of parameters, the rates
	corresponding to signal final states are higher for top-squark pair
	production via strong interaction than the case of electroweak
	associated production. 
	
	\subsection{Numerical scan}
	In order to identify the region of parameter space of our interest
	we perform an illustrative numerical scan of all relevant parameters.
	This scan is carried out using SUSPECT\cite{Djouadi:2002ze}
	to calculate the spectrum for a given
	set of input parameters, and then interfacing with 
	SUSYHIT\cite{Djouadi:2006bz} to obtain respective branching fractions of
	SUSY particle decays. Also micrOMEGAs\cite{Belanger:2005kh,Belanger:2004yn,Belanger:2006is,Belanger:2013oya} is interfaced for the calculation of DM related observables and
	then checking the constraints.
	
	We have set the 
	ranges of the most relevant parameters, including third generation soft squark masses ($\rm M_{Q_3},~M_{t_R} $), in the random scan
	(every unit is in GeV, wherever applicable):
	\br
	\rm  1.5 \leq {tan\beta} \leq 60, 30\leq M_1\leq 1000, 100\leq M_2\leq 3000,~~~~~~~~~~~~~~ \nonumber\\
	\rm 100\leq |\mu|\leq 1500,100\leq m_A\leq 1500, 
	600\leq M_{Q_3}\leq 2500, 600\leq M_{t_R}\leq 2500,
	\label{eq:nmssmpara}	
	\er
	while the other gaugino mass parameter is fixed as,
	\br
	\rm M_3=3~TeV.
	\er
	First two generations squark masses are assumed to be,
	\br
	\rm M_{Q_{1,2}}=3~TeV.
	\er
	The A-term corresponding to the third generation quark($\rm A_t$) plays an 
	important role in determining the lightest CP even SM-like Higgs boson mass, and
	it is varied in the range,
	\br
	\rm -6~TeV\leq A_t\leq 6~TeV.
	\er
	All the slepton masses of the first two generations are fixed to 2 TeV.	
	While performing the scan, each model point is tested with 
	PLANCK\cite{Aghanim:2018eyx} data (Eq.~\ref{eq:wmap}) and limits from 
	direct searches\cite{Akerib:2016vxi,Cui:2017nnn,Aprile:2018dbl,Agnes:2018ves,Agnese:2018col,Aprile:2019dbj,Amole:2019fdf,Adhikari:2019off,Ajaj:2019imk,Abdelhameed:2019hmk}. We focus only on the LSP of the mass range
	$\sim$50-500 GeV. 
	The presence of SM-like Higgs boson (h), with mass 
	125$\rm \pm 3 \; GeV$ is also ensured. Other absolute constraints 
	from LEP\cite{LEPSUSY}, for example, 
	$\rm m_{\rm\tilde{\chi}_1^{\pm}}\geq 103.5\; GeV$ 
	and $\rm m_{\rm H^{\pm}}> 78.6~GeV$ are imposed. 
	In addition,  Higgsbounds-5.5.0\cite{Bechtle:2008jh,Bechtle:2011sb,Bechtle:2013gu,Bechtle:2013wla,Bechtle:2015pma} is used to check the Higgs couplings and related
	measurements. The exclusion of top-squark-neutralino mass plane predicted by CMS\cite{Sirunyan:2019ctn,Sirunyan:2019xwh,Sirunyan:2019glc,CMS-PAS-SUS-19-011} and
	ATLAS\cite{Aad:2020sgw,ATLAS-CONF-2020-003,ATLAS-CONF-2020-046,Aaboud:2017aeu} experiments are also examined using the SModelS-1.2.3 package~\cite{Kraml:2013mwa,Ambrogi:2017neo}. Generally the SMS model
	with $\rm BR(\t\rightarrow\N_1 +t)=100\%$ is used to interpret data.
	Whereas, in our scenario, $\rm BR(\t\rightarrow\N_1 +t)\sim 10\%$ implies much weaker
	exclusion limits and consequently relatively light top-squarks ($\rm m_{\t}\sim 700~GeV$) are also found to be allowed.
	Performing the scan, Fig.\ref{fig:relic}(left)
	is plotted, where mainly the relic density and DD constraints are relaxed to show the effect of the compositions of $\rm \N_1$ on the relic density and DD measurements.
	
	Few representative benchmark points (BP) are chosen (see Table~\ref{tab:BPtable}), which are consistent with all constraints mentioned above. These BPs are used to obtain the signal sensitivities by performing the simulation of our proposed signal process. These BPs primarily represent two scenarios, namely ``mild-tempered neutralino'' and ``blind spots''. But under these broad pictures, they also encompass compressed and non-compressed spectrum corresponding to various choices of mass differences, $\rm \Delta m_1 = m_{\t} -(m_t + m_{\N_{2,3}})$ and $\rm \Delta m_2 = m_{\N_{2,3}} - m_{\N_1}$.
	\begin{table}[H]
		\centering
		\caption{\small Masses, branching fractions, DM observables for a few representative BPs and labeled those corresponding to the BS scenario. Energy units are in GeV, wherever applicable.}
		\resizebox{\textwidth}{!}{
\begin{tabular}{c c c c c c c c c c }
	\hline   & BP1 & BP2 & BP3 & BP4({\bf BS}) & BP5({\bf BS})& BP6 & BP7 & BP8({\bf BS}) & BP9({\bf BS}) \\ 
	\hline	$\rm M_1$  & 60.8 & 58.5 & 274.2  & 334.1 & 296.4 & 204.9 & 352.7 & 238.4 & 248.4\\
			$\rm M_2$  & 2784.4 & 2102.4 & 2719.2 & 1438.5 & 1494.1 & 1093.6 & 1860.2 & 1561.4 & 1071.0 \\
			$\rm \mu$  & 655.6 & 793.6  & 984.1  & -789.8  & -717.5 & -489.1 & -610.2 & -414.2 & -539.9\\
			$\rm m_A$  & 1252.7 & 953.2  & 584.1  & 712.7  & 585.6 & 453.9 & 762.1 & 459.3 & 543.8\\
			$\rm tan\beta$ & 7.5 & 6.0  & 6.6   & 6.1  & 6.2 & 5.0 & 5.8 & 6.3 & 6.7\\
			$\rm M_{Q_3}$  & 856.2 & 1102.2 & 2277.6& 1024.8 & 1544.2 & 770.1 & 824.4 & 765.8 & 811.5 \\
			$\rm M_{t_R}$  & 3552.0 & 1889  & 1688.8 & 2403.3  & 2061.9& 2381.8 & 2596.2 & 2088.9 & 2634.1 \\
	\hline  $\rm m_{\t}$ & 954 & 1059  & 1675  & 1038  & 1475& 688 & 804 & 635 & 765\\ 
			$\rm m_{\N_3}$  & 666 & 802 & 996  & 800 & 729& 494 & 620 & 424 & 550\\
			$\rm m_{\N_2}$ & 666 & 800 & 994  & 796 & 725& 499 & 618 & 422 & 545\\
			$\rm m_{\N_1}$   & 59 & 58  & 272  & 335  & 295& 207 & 354 & 238 & 249\\
			$\rm m_{\tilde{\chi}_1^\pm}$   & 664 & 799  & 993  & 795  & 725& 495 & 618 & 420 & 545\\
			$\rm m_{h}$  & 125 & 123   & 123  & 125 & 124& 123 & 123 & 124 & 125\\
			$\rm m_{H}$   & 1253 & 953 & 584  & 713 & 584& 454 & 763 & 460 & 544\\
	\hline  $\rm N_{11}^2$   & 0.995 & 0.996 & 0.996  & 0.996 & 0.996& 0.99 & 0.99 & 0.976 & 0.99\\
		    $\rm N_{13}^2+N_{14}^2$   & 0.005 & 0.003 & 0.003  & 0.004 & 0.004& 0.01 & 0.01 & 0.023 & 0.01\\		
	\hline  $\rm \Omega h^2$ & 0.129 & 0.122 & 0.119 & 0.112 & 0.121 & 0.110 & 0.117 & 0.119 &0.110\\			
		    $\rm \sigma_{SI}(10^{-11}\:pb)$ 	 &  $5.1$ & $5.2$ &  $10$ &  $0.009$ &  $0.02$ &  $1.9$ &  $2.4$ &  $0.69$ & $0.002$\\
		    $\rm \sigma_{SD}(p)(10^{-7}\:pb)$ &  $7.2$ & $3.2$ &  $1.6$ &  $5.2$ &  $7.5$ &  $32$ &  $21$ &  $100$ & $26$\\
		    $\rm \sigma_{SD}(n)(10^{-7}\:pb)$ &  $5.7$ & $2.5$ &  $1.3$ &  $4.1$ &  $5.8$ &  $25$ &  $12$ &  $78$ & $20$\\
	\hline	$\rm BR(\tilde{t}_1\rightarrow \tilde{\chi}_1^0 +\:t)$  & 0.05 & 0.11& 0.16  & 0.08 & 0.05& 0.11 & 0.15  & 0.08 & 0.08\\
			$\rm BR(\tilde{t}_1\rightarrow \tilde{\chi}_2^0 +\:t)$  & 0.49 & 0.31 & 0.20  & 0.37 & 0.34& 0.32 & 0.33& 0.34 & 0.33\\
			$\rm BR(\tilde{t}_1\rightarrow \tilde{\chi}_3^0 +\:t)$  & 0.42 & 0.51 & 0.22  & 0.49 & 0.43& 0.49 & 0.45& 0.52 & 0.50\\
			$\rm BR(\tilde{\chi}_2^0\rightarrow \tilde{\chi}_1^0 +\:h)$ & 0.70 & 0.73 & 0.70  & 0.73 & 0.72& 0.85 & 0.83 & 0.78 & 0.75\\
			$\rm BR(\tilde{\chi}_3^0\rightarrow \tilde{\chi}_1^0 +\:h)$ & 0.28  & 0.27 &0.12  & 0.24 & 0.25& 0.04 & 0.12 & 0.09 & 0.19\\
	\hline
\end{tabular}
}
		\label{tab:BPtable}
	\end{table}
	Notice also that for all cases of BPs, $\rm m_{\t}$ varies from 600-1700 GeV and for all cases $\rm BR(\t\rightarrow\N_1 +t)$ is subdominant,
	while $\rm \N_{2,3}\rightarrow \N_1 +h$ is dominant.
	Performing the simulation of signal and backgrounds,
	signal sensitivities are presented for all these BPs.
	
	\section{Signal and Background}
	
	As discussed before, we consider the following process where the lightest
	neutralino originates from the decay of second and third lightest neutralino($\N_{2,3}$)
	produced via top-squark production as shown below,
	\begin{figure}[H]
		\centering
		\includegraphics[width=9cm]{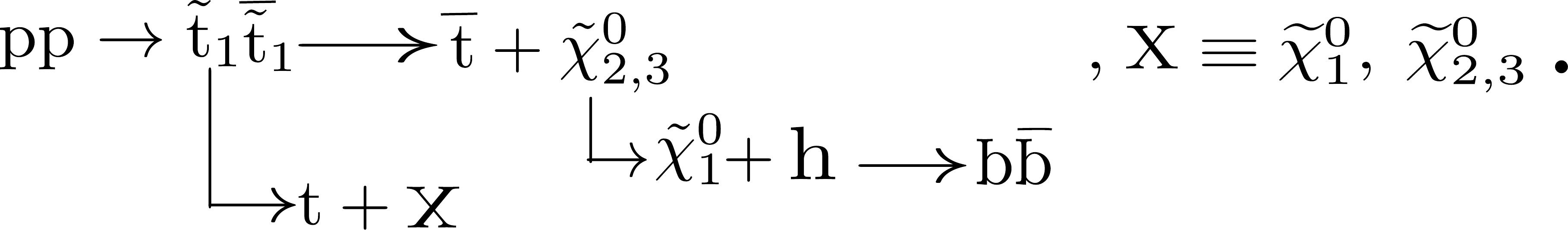}
		\label{fig:signal}
	\end{figure}
	Since in this scenario, the $\rm \N_2/\N_3$ are dominantly Higgsino-like,
	hence BR$\rm (\t \to t+ \N_{2,3})$ is larger than
	the BR($\rm \t \to t + \N_1$). Subsequently, the higher neutralino state (either $\rm \N_2$ or $\rm \N_3$) dominantly decays to SM-like Higgs boson and $\rm \N_1$.
	Here $\rm X \equiv \N_1, \N_{2,3}$ leads to 
	a $\rm \N_1$ accompanied by either Z or $\rm h$ in the final state.
	We focus only on single Higgs boson in the final state. 
	However, we found that
	the contribution of di-Higgs boson events in the
	signal is negligible.
	The $\rm b \bar b $ channel of Higgs boson decay is considered owing to its higher BR and comparatively easy to reconstruct its mass. The pair of lightest neutralinos escape the detector leading to a huge amount of missing energy in the final state. Moreover, there is another pair of b-jets originating from two top quarks. Hence, the final state of the signal event is characterized by,
	\br
	\rm h_{b\bar b}+ \ell + \MET+ (\geq 1) ~b-jets; \ \ \ell =e,\mu.
	\er
	We found that the contribution of di-Higgs production to the signal event is negligible. It is known that QCD is the main source of background corresponding to any pure hadronic final state. Hence, in order to eliminate it, the leptonic decay of one of the top quarks is considered. We require the presence of only one lepton in the final state. The other dominant SM backgrounds are:
	\br
	\rm p~p \rightarrow t\bar{t}(1\ell),~ t\bar{t}(2\ell),~ t\bar{t}h,~ t\bar{t}Z,~ t\bar{t}b\bar{b}
	\er
	where, the combination of two b's coming from the top, $\rm h$, Z, or gluon splitting mimics the signal b-jets from Higgs decay. The lepton and $\rm \MET$ arise from the semi-leptonic decay of one of the top quarks, while the other top decays hadronically.
	
	It is to be noted that in the signal events, the angular separation between two
	b-jets depends on the boost of Higgs boson,
	which is determined by the mass differences,
	$\rm \Delta m_1 = m_{\t} -(m_t + m_{\N_{2,3}})$ and $\rm \Delta m_2 = m_{\N_{2,3}} - m_{\N_1}$.
	Accordingly, we simulate signal events in resolved and non-resolved categories
	depending on the boost of Higgs boson. In Table~\ref{tab:BPtable}, BP1-BP5 correspond to the non-resolved category while BP6-BP9 represent the resolved one. For the boosted case,
	two b-jets likely to appear as a single fat jet, which we refer to as the ``Higgs jet(HJ)''
	now onwards.
	
	The PYTHIA8 \cite{Sjostrand:2006za,Sjostrand:2007gs} is used 
	to generate $\rm t\bar{t}(1\ell),~ t\bar{t}(2\ell)$ events, 
	while the other background processes are generated using 
	Madgraph5-aMC@NLO-2.7.3\cite{Alwall:2014hca} interfacing with PYTHIA8,
	for showering and hadronization. Signal events are generated in 
	Madgraph5-aMC@NLO-2.7.3 using UFO for the MSSM (MSSM-SLHA2), 
	where the parameter card is generated from SLHA file \cite{Skands:2003cj}, obtained from SUSYHIT, corresponding to each BP. 
	The same SLHA file is used for subsequent showering of signal events 
	in PYTHIA8. Detector effects are taken into account by passing all 
	signal and background events through Delphes-3.4.2\cite{deFavereau:2013fsa} 
	using the CMS detector card{\footnote{Results are checked with ATLAS card as well, and no appreciable change is observed.}}. 
	
	In the simulation, the following selections are imposed, where objects are selected using Delphes inputs.\\
	(1) {\bf Lepton selection :} Leptons are selected 
	with $\rm p_T>20~GeV$ and $|\eta|<2.5$.  Isolation is 
	ensured using mini-isolation criteria by checking e-flow objects of Delphes 
	as follows~\cite{Khachatryan:2016uwr}:
	\br
	\rm \frac{\sum p_{T}^{R<r}}{p_{T,\ell}}<I , \;\; \ell=e,\mu.
	\er
	Here $\rm r = \frac{10.0}{p_{T,\ell}}$ and $\rm I =0.12$ and 0.25 for e and $\mu$ respectively.
	
	(2) {\bf Missing transverse momentum ($\rm \MET$) :} The missing transverse 
	momentum is constructed  by taking the resultant momenta of all 
	visible particles and then reversing the direction, i.e. ${\vec{\rm p}_{\rm T}}= - \sum \vec{\rm p}_{\rm T}^{~i}$, 
	where $i$ runs over all constructed visible collection from the detector. A cut $\rm \MET>200(150)~GeV$ is imposed for events in the non-resolved(resolved) category.
	
	(3) {\bf HJ selection:} The reconstruction of HJ is performed in two ways depending on the boost of the Higgs boson, i.e., resolved and non-resolved categories, as described below.
	\begin{itemize}
		\item HJ in non-resolved category: At first, fat jets are constructed taking inputs from Delphes, using 
		Fastjet3.3.2\cite{Cacciari:2011ma} with 
		Cambridge-Aachen\cite{Dokshitzer:1997in} algorithm and R=1.0. 
		Minimum $\rm p_T$ of the fatjets is set to be 100 GeV. These fat jets are then passed through mass-drop Tagger (MDT)\cite{Butterworth:2008iy,Dasgupta:2013ihk} with $\mu$ =0.667 
		and $\rm y_{cut} >0.09$ to remove contamination due to soft radiation. The subjets of the `tagged fat jet' are 
		further matched with the b-quarks of the event which are selected within $|\eta|<2.5$ and with a matching cone $\rm \Delta R<0.3$. 
		When both the subjets are found to be b-like, we call the tagged 
		fatjet as the HJ ($\rm J_{bb}$). We also checked the presence of B-hadron in the b-like subjets and found that for about 95$\%$ cases, it exists.
		
		\item Resolved category: In this case, jets, subject to cuts $\rm p_{T}^{j}>$20 GeV and
		$|\eta|<$4.0, are constructed from e-flow objects of Delphes, using Fastjet3.3.2\cite{Cacciari:2011ma}, but with the Anti-$\rm k_T$\cite{Cacciari:2008gp} algorithm with a jet size parameter R=0.5. Using the same technique as above, by matching jets with b-quarks of the event, b-like jets are identified. The pair of b-like jets that construct the invariant mass closest to Higgs boson mass within the range 100 GeV$\rm \leq m_{HJ}\leq$150 GeV, is identified as HJ, and the resultant four-momentum of the two jets is regarded as the momentum of HJ.
	\end{itemize}
	
\begin{figure}[H]
		\begin{subfigure}[b]{0.48\textwidth}
			\centering
			\includegraphics[width=7.5 cm]{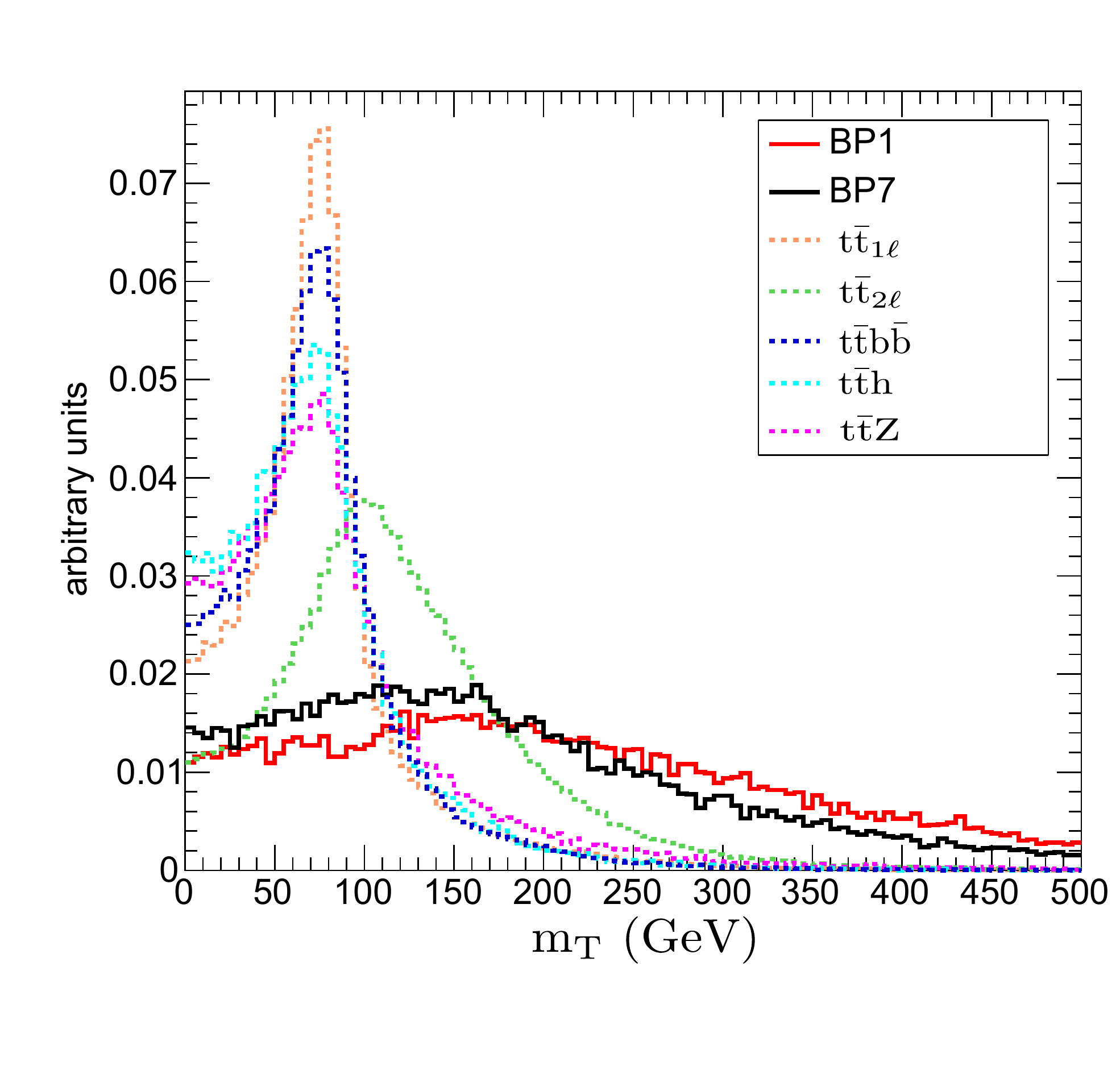}
		\end{subfigure}
		\begin{subfigure}[b]{0.48\textwidth}
			\centering
			\includegraphics[width=7.5 cm]{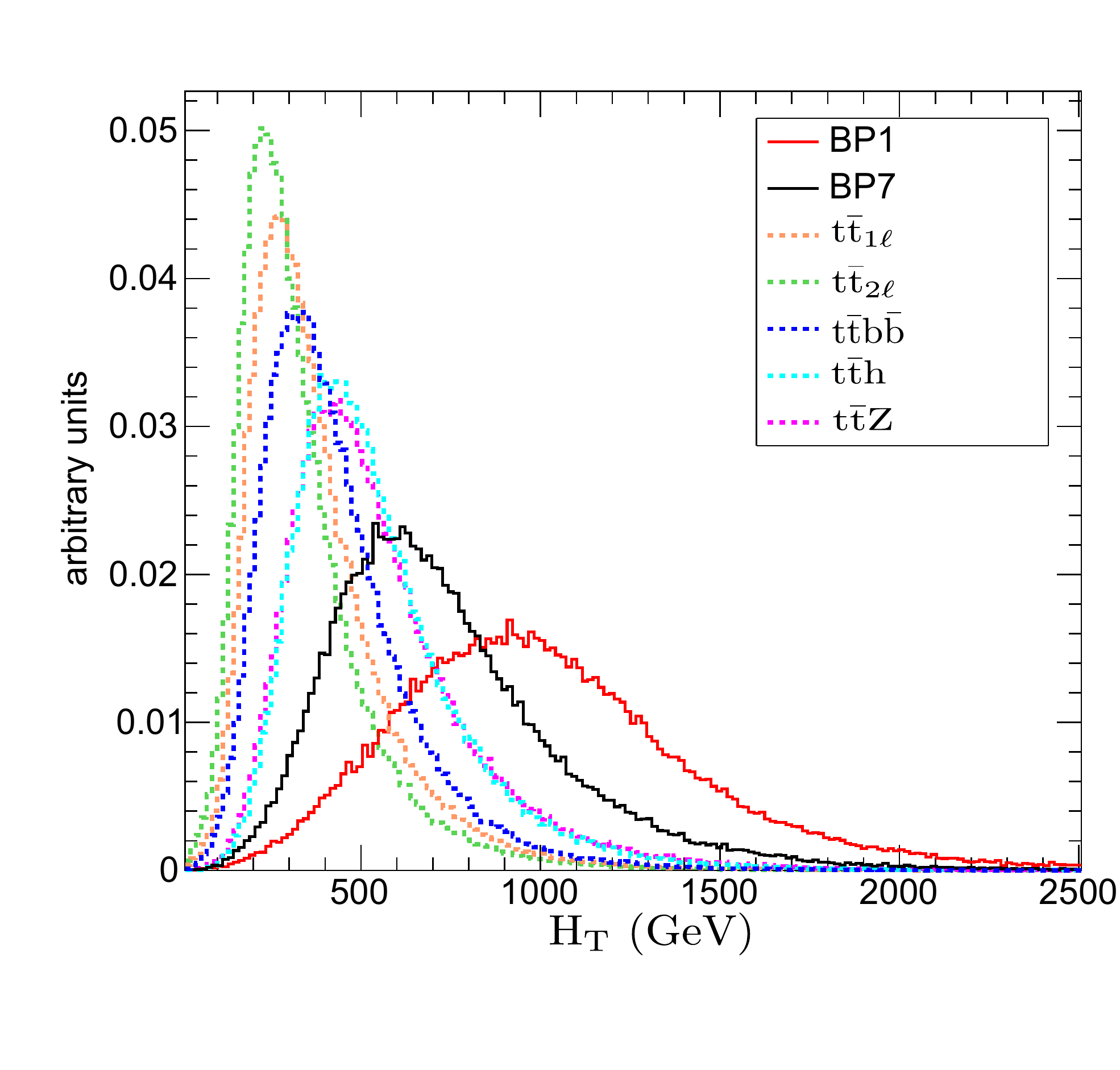}
		\end{subfigure}
		\caption{\small Transverse mass between (Eq.~\ref{eq:mtbb})lepton and $\rm \MET$ (left) and $\rm H_T$ (right) for BP1, BP7 and dominating backgrounds.}
		\label{fig:MTHT}
	\end{figure}

	(4) {\bf Other jets and b-jets:} This selection also differs according to two categories.
	\begin{itemize}
		\item Jets in non-resolved category: Once HJ is constructed, the remaining hadrons are used to construct regular QCD jets through Fastjet3.3.2 with Anti-$\rm k_T$ algorithm setting R=0.5. Out of these jets, b-like jets are identified by matching technique with the remaining set of b-quarks in the event, which are not part of $\rm J_{bb}$.
		
		\item Jets in resolved category: The two b-jets, which are found to be related to the HJ, are removed from the list of jets and b-jets, and this new list is used further.
	\end{itemize}
	Furthermore, to suppress backgrounds, we impose few more selection cuts. 
	For example, the transverse mass between lepton and $\rm \MET$, 
	defined as,
	\br
	\rm m_T(\ell, \MET) = \sqrt{2\times p_T^{\ell}\times \MET\times (1-\cos\phi(\ell, \MET))},
	\label{eq:mtbb}
	\er
	is restricted by $\rm M_W$ for all semileptonic $\rm t\bar{t}$ background events as seen in the $\rm m_T$ distribution presented in Fig.~\ref{fig:MTHT}(left) along with signal events corresponding to two BPs. On the contrary, for signal events, having 
	a large $\rm \MET$ due to neutralinos, which is also not correlated with the 
	lepton coming from $\rm t\bar{t}$ decay, is expected to have a more wide $\rm m_T$ distribution without any peaks (see Fig.~\ref{fig:MTHT} (left)). Hence a cut $\rm m_T(\ell,\MET)\geq110~GeV$ turns out to be very effective in eliminating a certain fraction of the background.
	Another discriminating variable is $\rm H_T$, defined as the scalar sum 
	of $\rm p_T$ of all jets except those that constitute HJ. For signal events, larger number of harder jets exist leading to higher $\rm H_T$ as seen 
	from the distribution shown in Fig.~\ref{fig:MTHT} (right). A cut $\rm H_T\geq 500~GeV$ turns out to be useful to reject background events substantially.
	\begin{figure}[H]
		\centering
		\includegraphics[width=8cm]{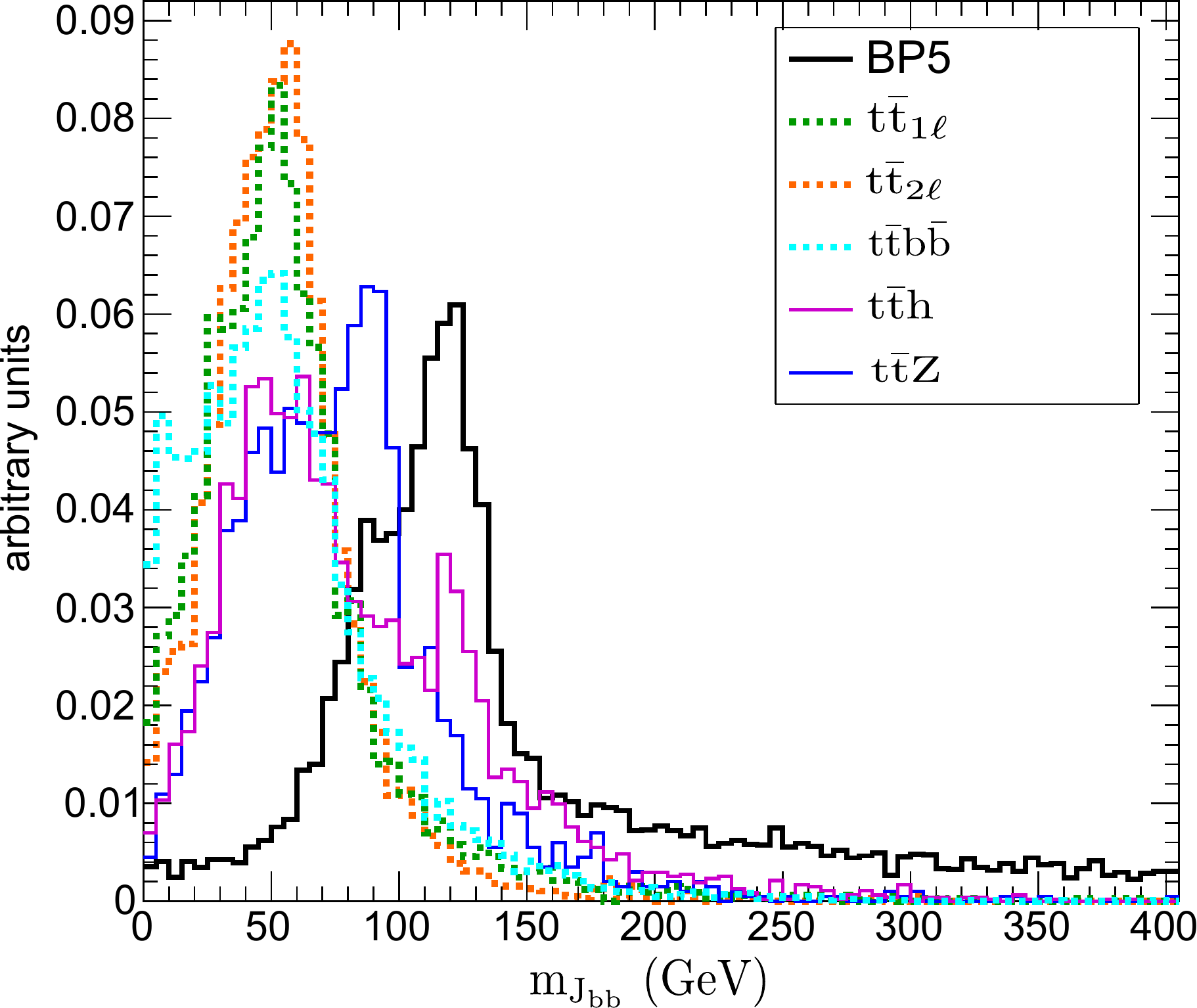}
		\caption{\small Reconstructed mass of Higgs Jet (HJ) for a representative signal(BP5) point and dominant backgrounds.}
		\label{fig:Mhiggs}
	\end{figure}
	In the case of the non-resolved category, the mass distribution of $\rm J_{bb}$
	shows a clear peak at $\sim$125 GeV, which is absent in most of the
	backgrounds, and very small for $\rm t\bar{t}h$
	as shown in  Fig.~\ref{fig:Mhiggs}. 
	Thus the selection of $\rm m_{J_{bb}}>100 ~GeV$
	is found to be useful in eliminating significant background events. In the resolved category case, this mass
	requirement is already imposed while constructing HJ. The presence
	of HJ with a specific mass requirement is a very important feature of our signal and helps to eliminate almost all the $\rm t\bar{t}$ backgrounds by enormous amount except $\rm t\bar{t}h$ process where the source of $\rm J_{bb}$ is same as the signal.

	Signal events are simulated for 9 BPs which are chosen in such a way that BP1-BP5 represent the non-resolved cases, whereas BP6-BP9 correspond to the resolved category. The BPs labeled as `BS' in the parenthesis correspond to BS scenario.
	In Table~\ref{tab:non-resolved}, the cross-section yields for the signal and background processes for the non-resolved categories are presented after imposing selection cuts. The first row presents the LO cross-sections of each processes, computed by Madgraph5-aMC@NLO-2.7.3, at the center of mass energy $\rm \sqrt{s}=13~TeV$, using NNPDF23LO \cite{Ball:2010de} 
	for parton distribution and choosing the dynamic QCD scale ($\rm Q^2=\sqrt{[m_{\t}^2+p_T^2(\t)][m_{\bar{\t}}^2+p_T^2(\bar{\t})]}$ ). Higher order effects are taken into account by multiplying respective K-factors($\rm K=\frac{\sigma_{NLO}}{\sigma_{LO}}$). A K-factor of 1.4 is used for top-squark pair production(for NNPDF31LO)\cite{Broggio:2013uba} and $\rm t \bar t$~\cite{Melnikov:2009dn,Kidonakis:2008mu}. Whereas, for $\rm t \bar t h$, $\rm t \bar t Z$ and $\rm t \bar t b\bar{b}$ K-factors are considered to be 1.2~\cite{Beenakker:2002nc}, 1.35~\cite{Kardos:2011na} and 1.8~\cite{Buccioni:2019plc} respectively. As indicated in the table, the $\rm m_{J_{bb}}$ cut is very useful to eliminate backgrounds significantly. In addition, the $\rm m_T$ cut also kills backgrounds substantially.
	\begin{table}[H]
		\caption{\small Cross-section (in fb) yields after each set of selection cuts for signal points in the non-resolved category and background processes.}
		\centering
		\resizebox{\textwidth}{!}{\begin{tabular}{ccccccccccc}
		\hline   & BP1  & BP2 & BP3 & BP4({\bf BS}) & BP5({\bf BS}) & $\rm t\bar{t} (1\ell)$  & $\rm t\bar{t} (2\ell)$  & $\rm t\bar{t} h$  & $\rm t\bar{t} Z$  & $\rm t\bar{t} b \bar{b}$  \\ 
		\hline Cross-secton(LO) (fb) &  6 & 3 & 0.06 & 3 & 0.18 & {$178500$}  & {$36000$}  & 400  & 584  & {$13700$}  \\ 
		
		$\rm \MET>200~ GeV$  & {$4.9$} & {$2.5$} & {$0.054$}  & {$2.4$} & {$0.17$} & {$2695$}  & {$592.5$}  & {$12.6$}  & {$38.8$}  & {$186.7$}  \\ 
		
		No. of $\rm \ell=1$  & {$1.5$} & {$0.73$} & {$0.02$} & {$0.7$} & {$0.05$} & {$1419$}  & {$291.2$}  & {$5.1$}  & {$11.5$}  & {$71.1$}  \\ 
		
		No. of $\rm J_{bb} =1$ & {$0.4$}  & {$0.2$} & {$0.004$} & {$0.2$} & {$0.014$} & {$33.8$}  & {$12.1$}  & {$1.1$} & {$0.8$} & {$10.6$}  \\ 
		
		$\rm m_{J_{bb}}>100$ GeV  & {$0.3$} & {$0.15$} & {$0.003$} & {$0.14$} & {$0.01$} & {$11.6$}  & {$2.7$}  & {$0.7$}  & {$0.3$}  & {$2.6$}  \\ 
		
		No. of b-jets$\geq 1$ &  {$0.15$} & {$0.07$} & {$0.001$} & {$0.07$} & {$0.0036$} & {$1.0$}  & {$0.25$}  & {$0.3$}  & {$0.1$}  & {$0.8$}  \\ 
		
		$\rm H_T>$500~GeV  & {$0.1$} & {$0.05$} & {$0.0008$} & {$0.05$} & {$0.003$} & {$0.25$}  & {$0.07$}  & {$0.1$}  & {$0.05$}  & {$0.1$}  \\ 
		
		$\rm m_{T}(\ell,\MET) \geq $110 GeV & {$0.08$} & {$0.043$} & {$0.0007$} & {$0.04$} & {$0.003$} & {$0.04$}  & {$0.07$}  & {$0.02$}  & {$0.006$}  & {$0.02$}  \\ 
		$\rm \sigma\times K$-factor & 0.12 & 0.06 & {$0.001$} & 0.056 & {$ 0.004$} & {$0.06$} & 0.1 & 0.024 & 0.008 & {$0.04$}\\
		\hline \end{tabular}
	}
		\label{tab:non-resolved}
	\end{table}
	\begin{table}[H]
		\caption{\small Same as in Table~\ref{tab:non-resolved}, but for resolved category.}
		\centering
		\resizebox{\textwidth}{!}{\begin{tabular}{cccccccccc}
 \hline   & BP6  & BP7 & BP8({\bf BS}) & BP9({\bf BS}) & $\rm t\bar{t} (1\ell)$  & $\rm t\bar{t} (2\ell)$  & $\rm t\bar{t} h$  & $\rm t\bar{t} Z$  & $\rm t\bar{t} b \bar{b}$  \\ 
 \hline Cross-secton (fb)  & 53  & 19 & 88 & 27 & {$178500$}  & {$36000$}  & 400  & 584  & 13700  \\ 
 $\rm \MET>150~ GeV$  & {$36.4$}  & {$14.1$} & {$51.2$} & {$20.5$} & {$8560$}  & {$2100$}  & {$31.3$}  & {$76.5$}  & {$555.6$}  \\ 
 No. of $\rm\ell=1$  & {$9.8$}  & {$3.8$}  & {$13.9$} & {$5.6$} & {$4364$}  & {$1050$}  & {$12.2$}  & {$23.8$}  & {$204.1$}  \\ 
 No. of $\rm J_{bb} =1$  & {$3.9$}  & {$1.5$}  & {$5.4$} & {$2.3$} & {$564.2$}  & {$145.9$}  & {$4.6$}  & {$3.9$}  & {$53.3$}  \\ 
 No. of b-jets$\geq 1$  & {$2.9$}  & {$1.1$} & {$4.0$} & {$1.8$} & {$49.5$}  & {$11.0$}  & {$3.6$}  & {$1.3$}  & {$35.7$}  \\ 
   	$\rm H_T$$>$500~GeV& {$1.7$}  & {$0.7$} & {$2.3$} & {$1.2$} & {$22.3$}  & {$3.6$}  & {$1.7$}  & {$0.6$}  & {$11.8$}  \\ 
   $\rm m_{T}(\ell,\MET) \geq $110 GeV& {$1.2$}  & {$0.5$} & {$1.5$} & {$0.9$} & {$5.3$}  & {$2.4$}  & {$0.4$}  & {$0.15$}  & {$2.6$}  \\ 
$\rm \sigma\times K$-factor & 1.7 & 0.7 & 2.1 & 1.22 & 7.5 & 3.36 & 0.43 & 0.20 & 4.7\\
\hline \end{tabular}
}
		\label{tab:resolved}
	\end{table}
	Similarly, cross section yields for the resolved category are presented in Table~\ref{tab:resolved}. It is clear that the selection of HJ, in this case, is not as efficient as the non-resolved category, but still having good discriminating power. In general, overall signal acceptance efficiency is 1-2$\%$; while for overall backgrounds, it is found to be 0.0001$\%$ for the non-resolved category and 0.007$\%$ for the resolved category. The total cross-sections of background events are found to be 0.232~fb for the non-resolved category and 16.2~fb for the resolved cases respectively. Finally, the signal sensitivities ($\rm \frac{S}{\sqrt{S+B}}$) are presented in Table~\ref{tab:sensitivity} for two high luminosity options $\rm \mathcal{L}=300~ fb^{-1}~ and ~3000~fb^{-1}$. It is to be noted that for the non-resolved category, the sensitivities are $\sim 2-3\sigma$ for $\mathcal{L}=300~\invfb$, whereas they are large($\sim 5-8\sigma$) for the resolved category, mainly because of the high production cross-sections, due to smaller top-squark masses. The tiny sensitivities for BP3 and BP5 can be attributed to a very low top-squark production cross-section because of its higher masses.  Assuming 10$\%$ background uncertainty the sensitivity for the BPs in resolved category drops by $\sim 7\%$ and for the non-resolved category, it reduces by about 0.1$\%$.
	
	\begin{table}[H]
		\caption{\small Signal significances($\rm \frac{S}{\sqrt{S+B}}$) for two luminosity options.}
		\centering
		\begin{tabular}{cccccc|cccc}
\hline 
& \multicolumn{5}{c|}{Non-resolved category} & \multicolumn{4}{c}{Resolved category}\\
\hline
 Luminosity $\rm (fb^{-1})$ & BP1 & BP2 & BP3 & BP4 & BP5 & BP6 & BP7 & BP8 & BP9\\ 
\hline 
$\rm 300$ 
& 3.5   &  2.0 & 0.035  & 1.8 & 0.14 & 7 & 2.9 & 8.5 & 5.0\\

$\rm 3000$
& 11  &  6 & 0.1  & 6 & 0.44 & 22  &  9 & 27 & 16\\
\hline

\end{tabular}
		\label{tab:sensitivity}
	\end{table}
	Though we obtain reasonable signal sensitivity in the resolved category, the acceptance efficiencies for backgrounds, in that case, are not appreciably small as in the non-resolved category in the cut-based method.
	In order to improve further, we carry out multivariate analysis (MVA) based on boosted decision tree (BDT) method within the framework of TMVA~\cite{Hocker:2007ht,Voss:2007jxm} framework.
	\subsection{Multivariate Analysis}

	The basic idea of MVA~\cite{Hocker:2007ht,Voss:2007jxm,Friedman:Statistics,Webb:Statistics,Kuncheva:Statistics} is to examine patterns in multidimensional
	data by considering several variables at once. Several kinematical variables are constructed, keeping in mind the features of signal events, for training purposes.
	Depending upon the performances of those variables, we use 13 of those for the non-resolved category and 15 for the resolved category to train signal and background samples. The description of those variables are presented in Tables~\ref{tab:rank_NR} and \ref{tab:rank_R} corresponding to BP5 for non-resolved category and BP7 for the resolved category respectively.
	\begin{table}[H]
		\centering\caption{\small Rank of variables in MVA for non-resolved category corresponding to BP5.}
		\scalebox{0.9}{	\begin{tabular}{c c c}
	 		\hline
	 		Rank & Variable  & Description\\
	 		\hline
	 		1 & $\rm m_h$  & Mass of $\rm J_{bb}$\\
	 		2  & HT      	&	Scalar sum of $\rm p_T$ of all jets outside $\rm J_{bb}$\\
	 		3 & $\rm \MET$     	&		Missing $\rm p_T$\\
	 		4 & $\rm \Delta R(\MET,J_{bb})$ &	$\rm \Delta R$ between $\rm \MET$ and $\rm J_{bb}$\\
	 		5 & $\rm p_{T}(J_{bb})$   &	$\rm p_T$ of $\rm J_{bb}$	\\ 		
	 		6 & $\rm p_T(\ell)$    &	$\rm p_T$ of leading lepton	\\
	 		7 & $\rm \Delta R(b_1,J_{bb})$ &  $\rm \Delta R$ between leading b-jet (outside $\rm J_{bb}$) and $\rm J_{bb}$\\
	 		8 & $\rm \Delta R(\MET,j)$ & $\rm \Delta R$ between $\rm \MET$ and leading jet outside $\rm J_{bb}$\\
	 		9 & $\rm Njets$        &	Number of jets outside $\rm J_{bb}$.	\\
	 		10 & $\rm M_{T}(\ell,\MET)$  &  Transverse mass of leading $\rm p_T$ lepton and $\rm \MET$ \\
	 		11 & $\rm N(\ell)$  & Number of leptons	\\
	 		12 & $\rm p_T$(b-jet)  & $\rm p_T$ of leading b-jet outside $\rm J_{bb}$  \\
	 		13 &  N(b-jet) 	&	Number of b-jets, outside $\rm J_{bb}$	\\
	 		\hline
	 	\end{tabular}
 	}
		\label{tab:rank_NR}
	\end{table} 
	The first column of these tables shows the ranking of these variables, which represents the relative importance in discriminating signal and backgrounds. The set of variables are the same for all BPs for a given category, but depending on the kinematics, the ranking of those variables is found to be little different.
	While doing MVA for each BP, overtraining tests are performed to ensure that there are no significant deviations between the performance of training and testing data.
	
	\begin{table}
		\centering
		\caption{\small Rank of variables in MVA for resolved category corresponding to BP7.}
		\scalebox{0.9}{\begin{tabular}{c  c c}
 		\hline
 		Rank & Variable  & Description\\
 		\hline
 		1  & $\rm \MET$     	&		Missing $\rm p_T$		\\
 		2  & $\rm m_h$  & Mass of $\rm J_{bb}$\\
 		3  & $\rm \Delta R(b_1,b_2)$  & $\rm \Delta R$ between two b-jets inside Higgs-jet.\\
 		4  & $\rm \Delta R(\MET,J_{bb})$ &	$\rm \Delta R$ between $\rm \MET$ and $\rm J_{bb}$\\
 		5  & $\rm p_{T}(b_1)/p_{T}(b_2)$ &pT ratio of two b-jets inside Higgs-jet. \\
 		6  & $\rm \Delta R(\MET,j)$ & $\rm \Delta R$ between $\rm \MET$ and leading jet\\

 		7  & $\rm \Delta R(b_1,J_{bb})$ &  $\rm \Delta R$ between leading b-jet (outside $\rm J_{bb}$) and $\rm J_{bb}$\\
 		8  & $\rm Njets$        &	Number of outside $\rm J_{bb}$.\\
 		9  & HT      	&	Scalar sum of $\rm p_T$ of all jets outside $\rm J_{bb}$\\

 		10  & $\rm p_T(\ell)$    &	$\rm p_T$ of leading lepton	\\
 		11  & $\rm M_{T}(\ell,\MET)$  &  Transverse mass of leading pT lepton and $\rm \MET$ \\
 		12  & $\rm p_{T}(J_{bb})$   &	$\rm p_T$ of $\rm J_{bb}$\\

 		13  &  N(b-jet) 	&	Number of b-jets, outside $\rm J_{bb}$\\
 		14  & $\rm N(\ell)$  & Number of leptons	\\
 		15 & $\rm p_T $(b-jet)  & $\rm p_T$ of leading b-jet outside $\rm J_{bb}$ \\
 		
 		\hline
 	\end{tabular}
}
		\label{tab:rank_R}
	\end{table}
	
	\begin{figure}[H]
		\begin{subfigure}[b]{0.49\textwidth}
			\centering
			\includegraphics[width=7.4cm]{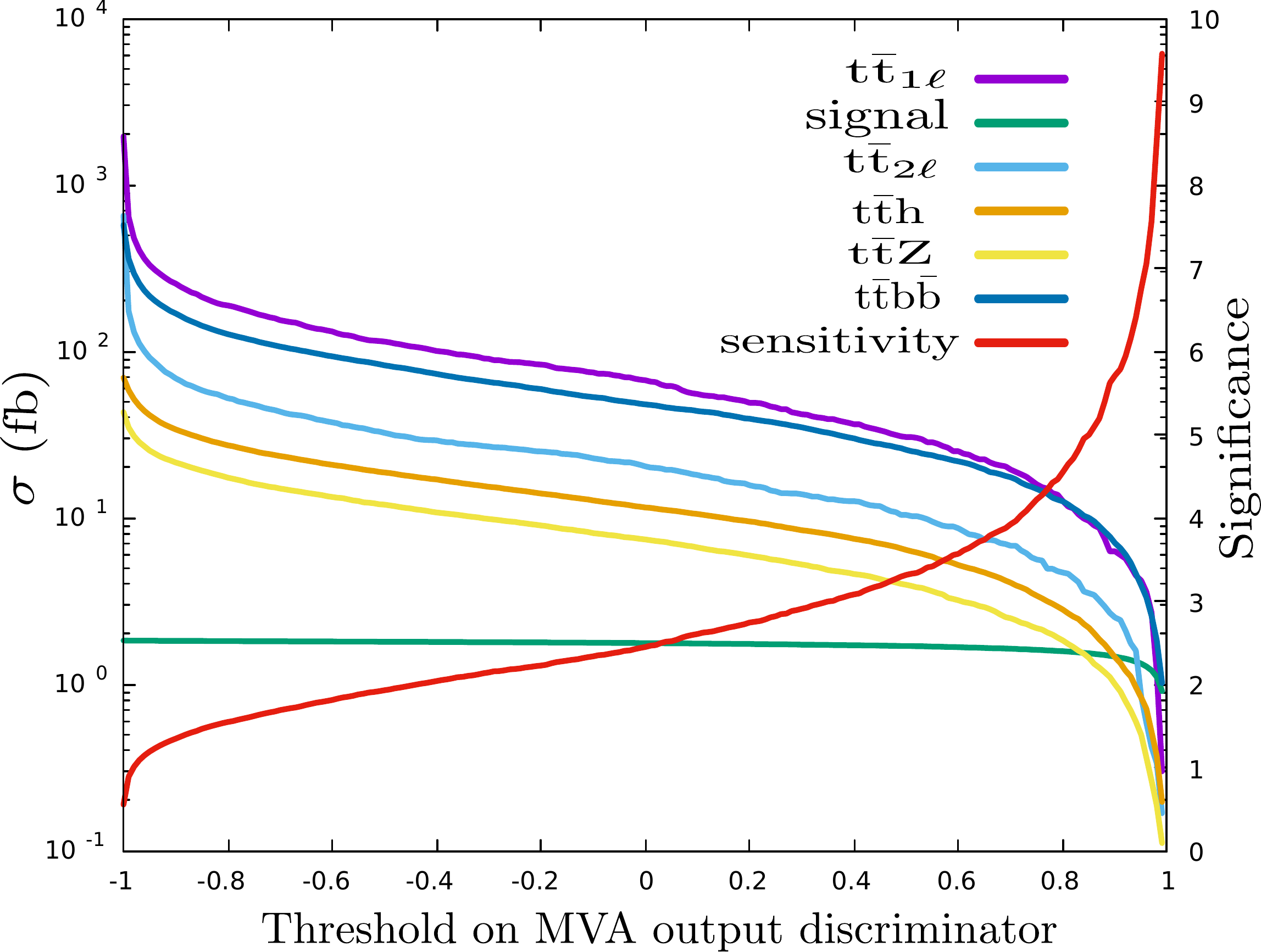}
		\end{subfigure}
		\begin{subfigure}[b]{0.49\textwidth}
			\centering
			\includegraphics[width=7.4 cm]{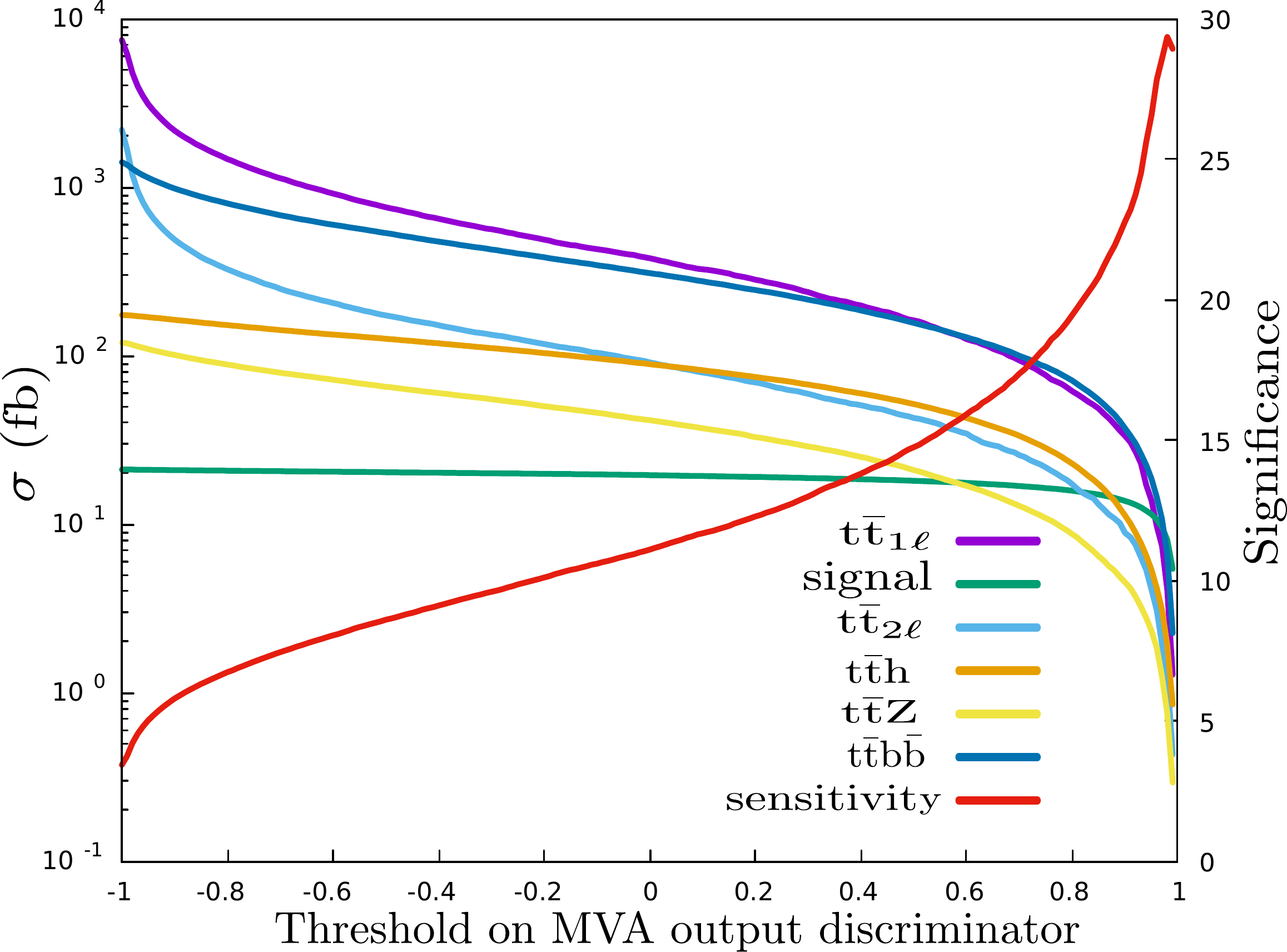}
		\end{subfigure}
		\caption{\small Signal and background yields as a function of threshold on MVA output discriminator along with the significance of signal corresponding to $\rm \mathcal{L}=300~fb^{-1}$ for signal point BP1 (left) and BP6(right).}
		\label{fig:MVA_809}
	\end{figure}
	In Fig.~\ref{fig:MVA_809}, the variation of cross section yields for signal and backgrounds and the signal significance ($\rm \frac{S}{\sqrt{S+B}}$) as a function of threshold on MVA output discriminator for luminosity $\rm \mathcal{L}=300~fb^{-1}$, is presented corresponding to the BP1 for the non-resolved category and BP6 for the resolved category case. It indicates that a sensitivity above $\sim 5\sigma$ can be achieved for luminosity $\rm \mathcal{L}=300~fb^{-1}$ corresponding to a cut of the classifier $>0.9$.

	Evidently, the achievable signal significance for all the BPs are presented in Table~\ref{tab:sens_mva} for two luminosity options. Clearly, the signal sensitivities are found to be well above $5\sigma$ at $\rm \cal{L}$$\rm = 300 ~fb^{-1}$, except for BP3 and BP5, where the production cross-section is too low due to a heavier top-squark mass.
	\begin{table}[H]
		\caption{\small Signal significances($\rm \frac{S}{\sqrt{S+B}}$) for two luminosity options applying MVA.}
		\centering
		\begin{tabular}{cccccc|cccc}
\hline  
& \multicolumn{5}{c|}{Non-resolved category} & \multicolumn{4}{c}{Resolved category}\\
\hline
 Luminosity $\rm (fb^{-1})$ & BP1 & BP2 & BP3 & BP4 & BP5 & BP6 & BP7 & BP8 & BP9  \\ 
\hline 
$\rm 300$ 
 & 6 &  4.5  & 0.14  & 3.6 & 0.35 & 24   &  9.5 & 27 & 15 \\

$\rm 3000$
 & 19 &  14 & 0.5  & 11 & 1.1& 75  &  30  & 85 & 47 \\
\hline

\end{tabular}
		\label{tab:sens_mva}
	\end{table}

	\section{Summary}
	In the MSSM framework, the lightest neutralino, an LSP of the mass $\sim\cal{O}$(100) GeV,  is found to be one of the best suitable DM candidates. However,
	the constraints from direct DM detection experiments and measurement of the relic density
	restrict the composition of the physical neutralino states. It is observed that, instead of a pure
	state, neutralino DM in MSSM is ``mild-tempered''
	where it is bino-dominated with a presence of little Higgsino, providing the best DM solution at this mass range.
	In this scenario, the DM annihilation process takes place via Higgs
	and gauge bosons where Higgsino content along with dominant
	bino helps to provide the
	right relic density. It is to be noted that, eventually the Higgsino composition in the LSP is strongly restricted by the limits of SI DM-nucleon scattering cross section measurements in the direct DM
	detection experiments, primarily by XENON1T. 
	Considering this DM solution, a numerical scan is performed to identify 
	the range of sensitive parameters, in particular, $\mu$ and $\rm M_1$ in the limit of a very large $\rm M_2$ value.   
	It is found that, with $\rm|\mu|-M_1 > M_Z$, the most preferred ranges
	are $\rm M_1 \sim 50 - 600$~GeV and $\rm \mu \sim 400- 1000$~GeV.  
	Moreover, there is a region of parameter space that is blind to
	the SI scattering cross section due to the interplay of parameters
	and cancellation among various amplitudes
	mediated by the lighter and heavier Higgs bosons. Consequently, in such cases, the
	Higgsino content in the lightest neutralino is not severely constrained. 
	In mild-tempered DM scenario, $\rm \N_1$ is
	accompanied with Higgsino-like $\N_{2,3}$ and $\widetilde{\chi}_1^{\pm}$
	having masses around $\mu$. It is indeed the case even for the region of
	parameters corresponding to the BS scenario.  
	Due to the gaugino-Higgsino-Higgs type of coupling,
	$\rm \N_{2,3}\rightarrow h+ \N_1$ decay rate gets enhanced,
	leading to an interesting phenomenology at the LHC corresponding to our considered scenario.
	
	We focus on the top-squark pair production to explore the mild-tempered neutralino scenario at the LHC. As $\rm BR(\t\rightarrow\N_1 +t)$ is very small, $\rm \N_1$ is indirectly produced through the production of $\N_{2,3}$.
	The presence of SM-Higgs boson in the final state adds an extra advantage to
	probe this channel.  Interestingly, this channel also provides an opportunity to probe the BS scenario. The signal is characterized by one HJ consisting of b-like jets or subjets,
	large $\rm \MET$, one lepton, plus at least one extra b-like jet.
	The HJ tagging turns out to be very efficient to separate out the signal
	from the debris of backgrounds. The presence of HJ adds robustness to this signal.
	
	Signal significances are presented for few illustrative BPs including BS scenario. We observe that for top-squarks of the mass range 600-1700 GeV, for most of the BPs, a reasonable signal sensitivity($\sim3-5\sigma$) can be achieved corresponding to $\rm \cal{L}$ $\rm $ = 300 ~fb$^{-1}$ luminosity option,  which goes up roughly by a factor of three for $\rm \cal{L}$$\rm = 3000 ~fb^{-1}$. Furthermore, we demonstrate that the sensitivities can be increased by employing
	MVA technique. Remarkably, we notice that, for the above luminosity options, and in particular for the resolved category case, the improvement is significant, by a factor of $\sim$3-4. The signal is detectable even for $\rm \cal{L}$ $\rm $ = 300 ~fb$^{-1}$ option except for BP3 and BP5 for which top-squark masses are $\rm \sim1.5~TeV$. For the center of mass energy $\rm \sqrt{s}=14~TeV$, which is the energy option for RUN3 experiment at the LHC, our projected sensitivities are expected to increase by 15-20$\%$ depending on the top-squark masses. A 10$\%$ uncertainty in background estimation reduces sensitivity by about 7$\%$ and $0.1\%$ for resolved and non-resolved category respectively.
	Our analysis shows that both the ``mild-tempered" neutralino providing a DM candidate in the framework of
	MSSM and also the BS scenario where the direct search is not sensitive,
	can be detected at the LHC with a reasonable sensitivity for projected luminosity
	options. \\

{\bf Acknowledgements}

The authors are thankful to Shivani Lomte, at affiliation of IISER, Pune, (now at University of Wisconsin) for collaborating on this project at an earlier stage. One of the authors, A.R, is thankful to Soham Bhattacharya for useful discussions and suggestions.

\bibliographystyle{utphys.bst}
\bibliography{paper.bib}

\end{document}